\documentclass[conference]{IEEEtran}
\IEEEoverridecommandlockouts
\usepackage{cite}
\usepackage{amsmath,amssymb,amsfonts}
\usepackage{algorithmic}
\usepackage{graphicx}
\usepackage{textcomp}
\usepackage{xcolor}
\usepackage{subfig}
\usepackage{graphicx}%
\usepackage{enumitem}
\usepackage{multirow}%
\usepackage{tabularx}
\usepackage{balance}
\usepackage{graphics}
\usepackage{url} 

\newif\ifhighlight
\highlighttrue 

\newcommand{\highlight}[1]{\ifhighlight{#1}\else#1\fi}

\def\BibTeX{{\rm B\kern-.05em{\sc i\kern-.025em b}\kern-.08em
    T\kern-.1667em\lower.7ex\hbox{E}\kern-.125emX}}
\begin{document}

\title{Social Robots as Social Proxies for Fostering Connection and
Empathy Towards Humanity\\
}


\author{
\IEEEauthorblockN{Jocelyn Shen}
\IEEEauthorblockA{\textit{MIT Media Lab} \\
Cambridge, USA \\
joceshen@mit.edu{   }}
\and
\IEEEauthorblockN{Audrey Lee}
\IEEEauthorblockA{\textit{MIT} \\
Cambridge, USA \\
audrey16@mit.edu}
\and
\IEEEauthorblockN{Sharifa Alghowinem}
\IEEEauthorblockA{\textit{MIT Media Lab} \\
Cambridge, USA \\
sharifah@mit.edu}
\and
\IEEEauthorblockN{River Adkins}
\IEEEauthorblockA{\textit{MIT} \\
Cambridge, USA \\
radkins@mit.edu}
\and
\IEEEauthorblockN{Cynthia Breazeal}
\IEEEauthorblockA{\textit{MIT Media Lab} \\
Cambridge, USA \\
breazeal@mit.edu}
\and
\IEEEauthorblockN{Hae Won Park}
\IEEEauthorblockA{\textit{MIT Media Lab} \\
Cambridge, USA \\
haewon@mit.edu}
}

\maketitle

\begin{abstract}
Despite living in an increasingly connected world, social isolation is a prevalent issue today. While social robots have been explored as tools to enhance social connection through companionship, their potential as asynchronous social platforms for fostering connection towards humanity has received less attention. In this work, we introduce \highlight{the design of a} social support companion that facilitates the exchange of emotionally relevant stories and scaffolds reflection to enhance feelings of connection \highlight{via five design dimensions.} We investigate how social robots can serve as ``social proxies'' facilitating human stories, passing stories from other human narrators to the user. To this end, we conduct a real-world deployment of 40 robot stations in users' homes \highlight{over the course of two weeks. Through thematic analysis of user interviews, we find that social proxy robots can foster connection towards other people's experiences via mechanisms such as identifying connections across stories or offering diverse perspectives. We present design guidelines from our study insights on the use of social robot systems that serve as social platforms to enhance human empathy and connection.}

\end{abstract}


\section{Introduction}
Despite living in an increasingly connected world, loneliness and apathy are widespread ~\cite{oviatt_technology_2021, buecker_is_2021}. Social isolation is fiercely detrimental to mental health and emotional resilience, but beyond mental health alone, social connections imbue the agency to take prosocial actions in the world \cite{jang_design_2023, morelli_emerging_2015, seppala_social_2013}. Motivated by this human need, many prior works have explored using embodied social robots as tools to improve social connection by offering companionship \cite{eyssel_loneliness_2013, pirhonen_can_2020, berridge_companion_2023}. Social robots offer many promises with regard to addressing social needs through their affective and personalized interaction, physical presence, and social reasoning capabilities \cite{nunez_effect_2019}. However, few have explored the role that social agents can serve as asynchronous social platforms that mediate empathetic exchanges and improve connection towards other people rather than enhancing the human-robot relationship \cite{narain_promoting_2020, rahwan_intelligent_2020, alves-oliveira_robots_nodate}. 

\begin{figure}[t]
  \includegraphics[width=\linewidth]{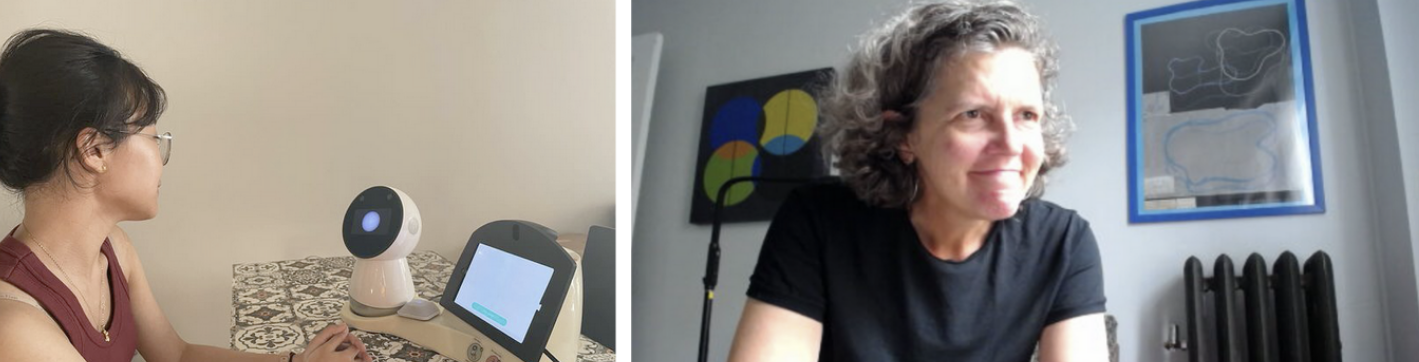}
  \caption{We deployed 40 in-home social robots as social platforms connecting the stories told by participants to other peoples' experiences to understand how robot story sharing method affects human connection and empathy.
  }
  \label{fig:teaser}
  \vspace{-15pt}
\end{figure}

In this work, we present \highlight{the design of }a novel one-on-one social support companion that fosters an exchange of emotionally relevant stories \highlight{and narrative therapy based reflection in order to bolster feelings of human connection in users. Our design introduces the concept of a ``\textbf{social proxy}'' agent, which we define as an agent that connects experiences across people} (Fig.~\ref{fig:teaser}). 
In this work, we focus on \highlight{understanding how to design social robot interactions for supporting social connectedness. } We find that in prior empathy related 1:1 human-robot interaction works, the robot's aim is typically to bond with the user \cite{ahmed_robots_2022, anzabi_exploring_2023, rosenthal-vonderputten_experimental_2013, spitale_socially_2022}. While human-robot bonding can lead to improved downstream social outcomes, social robots' roles in mediating human connection are under-explored. 



Through a real-world deployment of our system in people's homes, our work aims to answer the overall research question of, \textbf{can we use social robots as social proxies to foster empathy and connection towards other people?} Our overarching research question aims to trace the path between our proposed social robot interaction design, how the interaction influences user empathy and social connection, and how these effects \highlight{are related to explicit design affordances created by the social robot}. To this end, we answer the following:
    
\begin{itemize}
    \item \textbf{[RQ1]: \highlight{What is the impact of a social proxy robot sharing empathically resonant stories for fostering social connection and empathy towards people's stories? }
    }
      \vspace{-10pt}
    \item \textbf{[RQ2]: How is the effect of our social robot interaction \highlight{related to the} design features of the robot interaction?} 
\end{itemize}

\section{Related Work}
In this section, we discuss related work across social psychology, human-computer interaction (HCI), and human-robot interaction (HRI) on (1) how artificial agents, \highlight{and more broadly, computer-mediated communication has be used to foster connection, and (2) how design of story sharing interactions affects empathy and connection.}

\subsection{Technology-Mediated Interaction for Social Connection}
\highlight{While we introduce the design of ``social proxy'' agents, the conceptualization of this design paradigm is rooted in the fields of computer-mediated and AI-mediated communication \cite{hancock_ai-mediated_2020} where systems serve as intermediaries to improve communication between individuals or groups. For example, within HCI and computer-supported collaborative work (CSCW) communities, systems employ various strategies such as augmented/virtual realities (AR/VR) \cite{leong_social_2023-1, herrera_building_2018, oh_virtually_2016, mulvaney_social_2024}, online spaces \cite{poon_computer-mediated_2023}, ambient displays \cite{kucera_bedtime_2021, davis_promoting_2016}, and sharing biosignals and touch \cite{liu_significant_2021, wei_mediated_2023, k_miller_meeting_2021} for mediating human connection. Surveys of technologically mediated communication identify that design of such systems specifically aid in facilitating, inviting, and encouraging social interaction among people \cite{olsson_technologies_2020, stepanova_strategies_2022-1}.}

However, few works have explored how to design social robots as social proxies, despite the fact that social robots offer many advantageous qualities that make them suitable as social platforms to support connection towards humanity. One work by Narain et al.~\cite{narain_promoting_2020} uses a chatbot to facilitate positive message sharing between members of a social group but does not explore embodiment or how the user's personal experiences empathically relate to others. Other works use embodiment to channel the physical presence of a social partner \cite{nunez_effect_2019} or the sharing of living noise through a social robot to enhance social connectedness \cite{jeong_fribo_2018-2}. Social robots have also been used as social catalysts \cite{rahwan_intelligent_2020}, for example to promote inclusivity in synchronous group settings \cite{sebo_robots_2020, strohkorb_sebo_strategies_2020, weisswange_what_2023}. In contrast to these prior works, our work evaluates ways social robots can be extended to the domain of asynchronous social support, offering a new and personal modality to connect with others like that of social media. Our analysis traces how the social robot design leads to feelings of connection.

The rich modalities, both verbal and nonverbal cues, that embodied social robots possess allow them to express social information in more natural and impactful ways \cite{breazeal_effects_2005,wainer_role_2006,deng_embodiment_2019}. Prior works report that people empathize with and feel empathized by embodied agents specifically \cite{ventre-dominey_embodiment_2019}. Compared to virtual agents or chatbots, embodied conversation provides a greater sense of presence, which is most suitable for a study on social connection \cite{deng_embodiment_2019}. Furthermore, previous works have shown that people are often more comfortable disclosing sensitive personal information to virtual and physical AI agents than to other people \cite{bethel_using_2016-1,pickard_revealing_2016,lucas_its_2014, bethel_secret-sharing_2011}. 
Despite these characteristics, few works have explored how to use embodied social agents as social platforms for human connection as opposed to improving the bond between the agent and its user. 

\subsection{Delivering Empathic Stories}

\label{sec:deliverying_empathic_stories}
Our system is rooted in the fact that humans naturally share stories to emphasize close connections, and people resonate more with others when they self-disclose vulnerable stories \cite{keen_theory_2022, hodges_giving_2010}. Studies found that when the robot adopts self-disclosure as an interaction strategy, people would also disclose more to the robot and perceive the robot as more companion-like ~\cite{jeong_robotic_2023, laban_sharing_2023, westlund_measuring_2018,kory2019exploring}. Other works have found that when a social robot tells stories in the first person, people resonate more with the robot and report higher empathy and perceived likeability than stories told in third-person \cite{spitale_socially_2022}. \highlight{However, these studies that compare first vs third person story delivery typically refer to narrative voice (“I am sad” vs “The robot was sad”), whereas our concept of a social proxy is focused on the robot passing stories from others who can relate (e.g., “This is a story from a person who can relate to sadness …”). }

While much literature points towards the benefit of self disclosure and first-person storytelling, they focus primarily on studying the bonding and relationship building between \textit{human and robot}. In our work, we attempt to study the role a robot agent can play in fostering broader \textit{human} connection and compassionate love for humanity through sharing resonant stories, \highlight{i.e., we introduce a new paradigm of using robots as \textbf{social proxies} for \textbf{human connection}. We define the robot's proxy role for social connection as connecting the interactants to stories of other people who may relate and shares, the story on the narrator's behalf.  Through qualitative analysis of a real-world deployment study, we assess the intertwining relationship among robot design, users' perceptions toward the robot, and social connection. Our proposed investigation offers new design insights and a novel paradigm for thinking about human connection via social proxy agents.}

\begin{figure*}[t!]
    \centering
  \includegraphics[width=.75\linewidth]{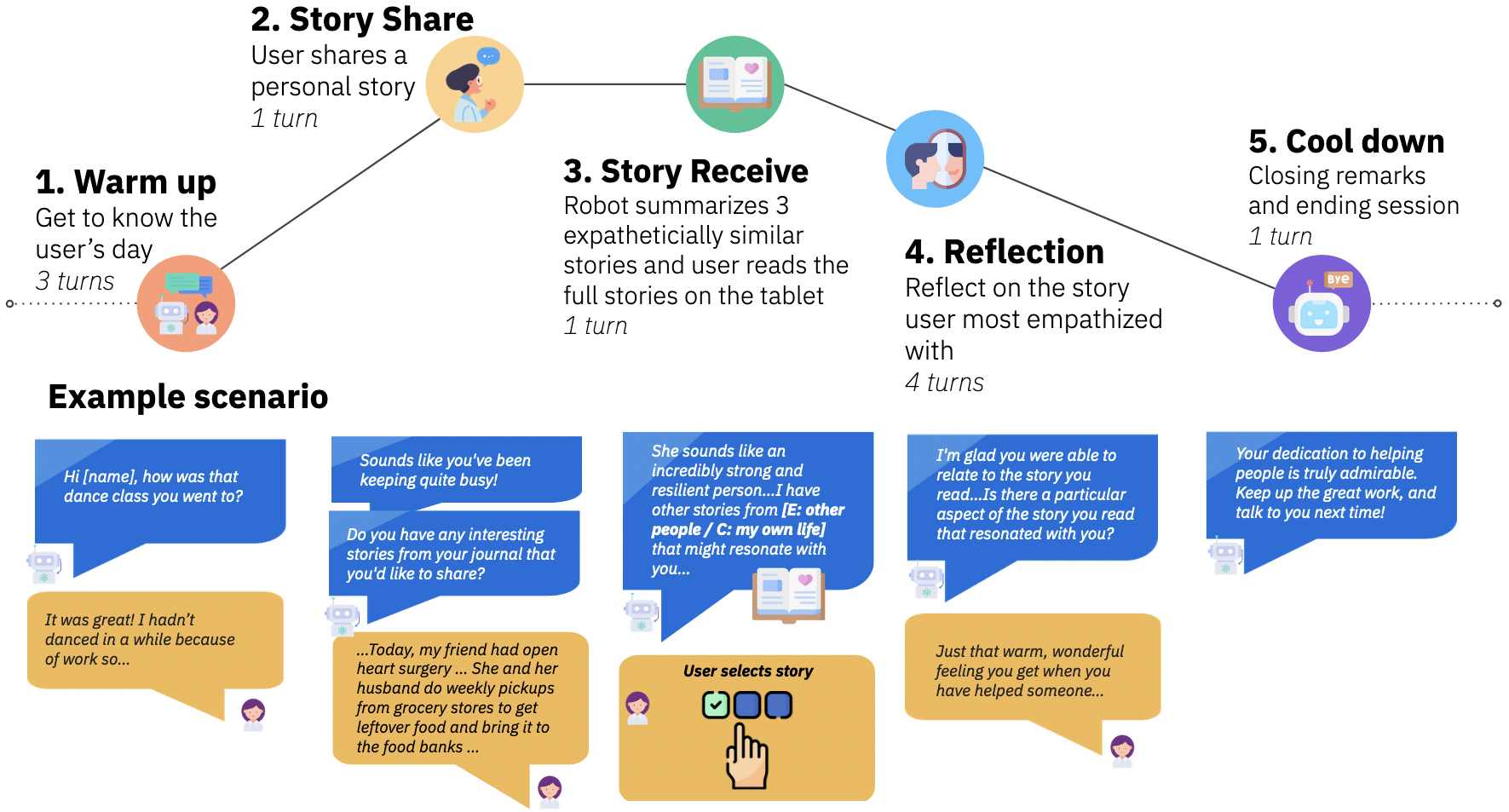}
  \caption{Interaction design phases, with examples from a participants' session shown in blue.}
  
  \label{fig:phases}
  \vspace{-10pt}
\end{figure*}

\section{Interaction Design}
We designed a companion-like social robot platform that fosters empathetic story sharing through personal journaling interactions, where the user shares their own story and the robot retrieves emotionally resonant stories before offering a narrative therapy based reflection of the stories. Note that while our interaction design was rooted in narrative therapy prompts, the design was \textit{not} intended to be used as a therapist, specifically for addressing mental health concerns. Rather, our system aims to cultivate social connectedness and empathy towards stories from the general population, and this distinction was made clear to all participants of our study prior to consent.


\subsection{Design Principles} \label{designprinciples}
We enumerate 5 design principles grounded in HRI/HCI and psychology literature that drove our interaction design, \highlight{as well as how these principals tie specifically to social proxy robots. }

\begin{itemize}[left=0pt]
    \item \highlight{\textbf{[D1] Physical embodiment of stories for greater emotional involvement}. 
    Prior works find that people exhibit more empathy towards physically embodied than disembodied robots \cite{kwak_what_2013}. For example, physical embodiment can improve attention to stories told by robots \cite{costa_emotional_2018}, which can lead to greater narrative transportation, or involvement into the world of a story. Transportation is known to lead to further empathy with narrators  \cite{johnson_transportation_2012}. Though virtually embodied agents and other computer-mediated communicative devices have studied proxy agents for connection \cite{narain_promoting_2020}, prior evidence regarding positive effects of physical embodiment motivates us to explore this new design paradigm as a physically embodied agent for supporting connection \cite{ventre-dominey_embodiment_2019}.   }
    \item \highlight{\textbf{[D2] Increasing self-disclosure via non-judging agents}.  Increased self disclosure has been shown to lead to more authentic conversations and thus deeper social connections in various settings \cite{utz_function_2015, laban_opening_2023, laban_sharing_2023}. Prior works indicate that people can feel more comfortable self-disclosing sensitive information with agents than other people \cite{bethel_using_2016-1, bethel_secret-sharing_2011, lucas_its_2014, noguchi_how_2023}, citing that participants felt ``less judged'' when disclosing to an agent. Therefore, it motivates developing research where AI agents are used as non-judging listening partners and social proxy agents for transmitting stories shared by real people instead of the robot itself. To enable the robot's role as a social proxy in our study, the stories users shared are matched based on empathetic relevance \cite{shen_modeling_2023}, and as such, the robot proxy could encourage user's self-disclosure by selecting authentic, emotionally relevant stories. }
    \item \highlight{\textbf{[D3] Long-term relational interaction}. Previous works in HRI indicate that personalization over long-term interactions with repeated encounters is crucial in the design of relational socially intelligent AI \cite{kory-westlund_long-term_2022}. Prior works indicate through long-term studies that participants can, over time, feel similar levels of social support from an empathetic social robot as they do with human peers \cite{ leite_long-term_2012}. While these works demonstrate that long-term interaction is important for relationship with the \textit{agent} \cite{kasap_building_2012}, we hypothesize that it is a necessary design feature for social proxy agents as well, as longitudinal relationship with the proxy can affect reception of the stories shared by others. As such, our design utilizes long-term memory, referring to shared experiences between the human and robot over time. }
    \item \highlight{\textbf{[D4] Adaptive and socially contextualized conversation}. In social psychology, lower level forms of adapting to social partners, such as mimicry and linguistic style matching are well studied \cite{niederhoffer_linguistic_2002}, and humans are also able to perform higher levels of social intelligence such as understanding the current context of the user's dialogue and responding appropriately. A social proxy agent that aims to enhance connection towards others must (1) respond appropriately to the personal experience shared by the user so their experience is acknowledged as grounded in theories of reciprocity and social exchange \cite{gouldner_norm_1960, emerson_social_1976}, and (2) understand and relate themes dynamically to emphasize connections between the user and the target's stories~\cite{bhattacharjee_i_2022, pillemer_remembering_1992, shen_modeling_2023, dinakar_you_2012} instead of solely leaving such interpretation to the user. To this end, our interaction utilizes a database with thousands of stories to match the diversity of users, and we add a reflection phase in the interaction where the robot enables multi-turn conversation about the story with a third-person perspective.}
    \item \highlight{\textbf{[D5] Clarify story ownership for narrative authenticity}. Prior works have shown that humans prefer to empathize with human vs AI targets in stories, but are more willing to empathize with AI targets when authorship is transparent \cite{shen_empathy_2024, giorgi_2023, laban_opening_2023}. As such, the main design principal of social proxy agents is clarifying that stories are shared \textit{through} the robot, but are authored by human narrators. This design choice especially differentiates the \textit{social proxy} from the existing paradigm of \textit{social agents}, where robots self disclose their \textit{own} experiences \cite{spitale_socially_2022, tsumura_influence_2023}. This enables the focus to be on facilitating social connectedness between people. Since it is an understudied design paradigm, our study design includes investigating the social proxy design further compared to the social agent design.}

\end{itemize}



\subsection{Interaction Phases and Conversation Design}
From our guiding design principles, we then developed the following interaction phases (Fig.~\ref{fig:phases}):

\begin{figure*}[t!]%
    \centering
    \subfloat[\centering Data storage, real-time sensors, and APIs]{{\includegraphics[width=0.4\textwidth]{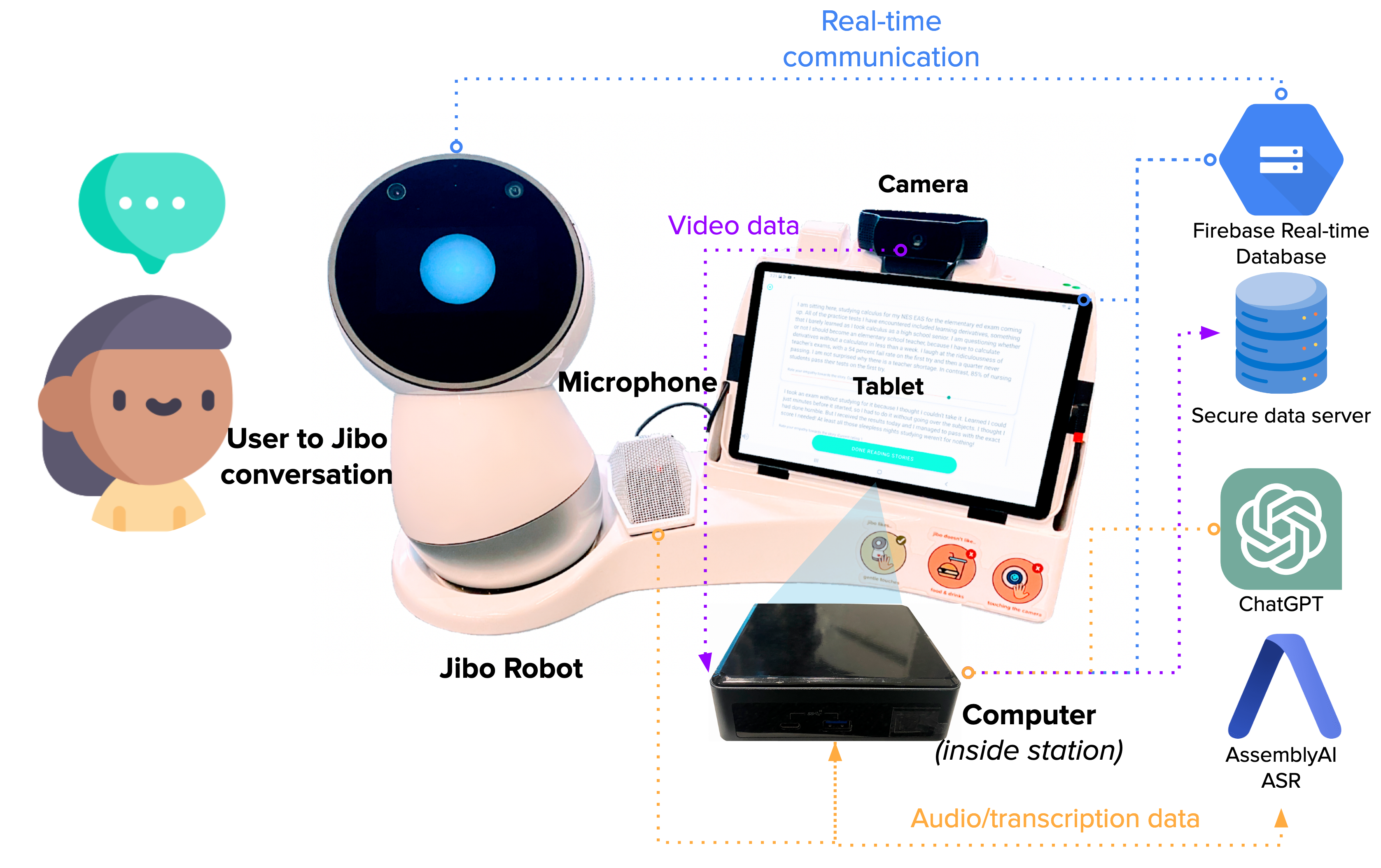} }}%
    \qquad
    \subfloat[\centering Tablet feed displaying stories and empathy sliders]{{\includegraphics[width=0.35\textwidth]{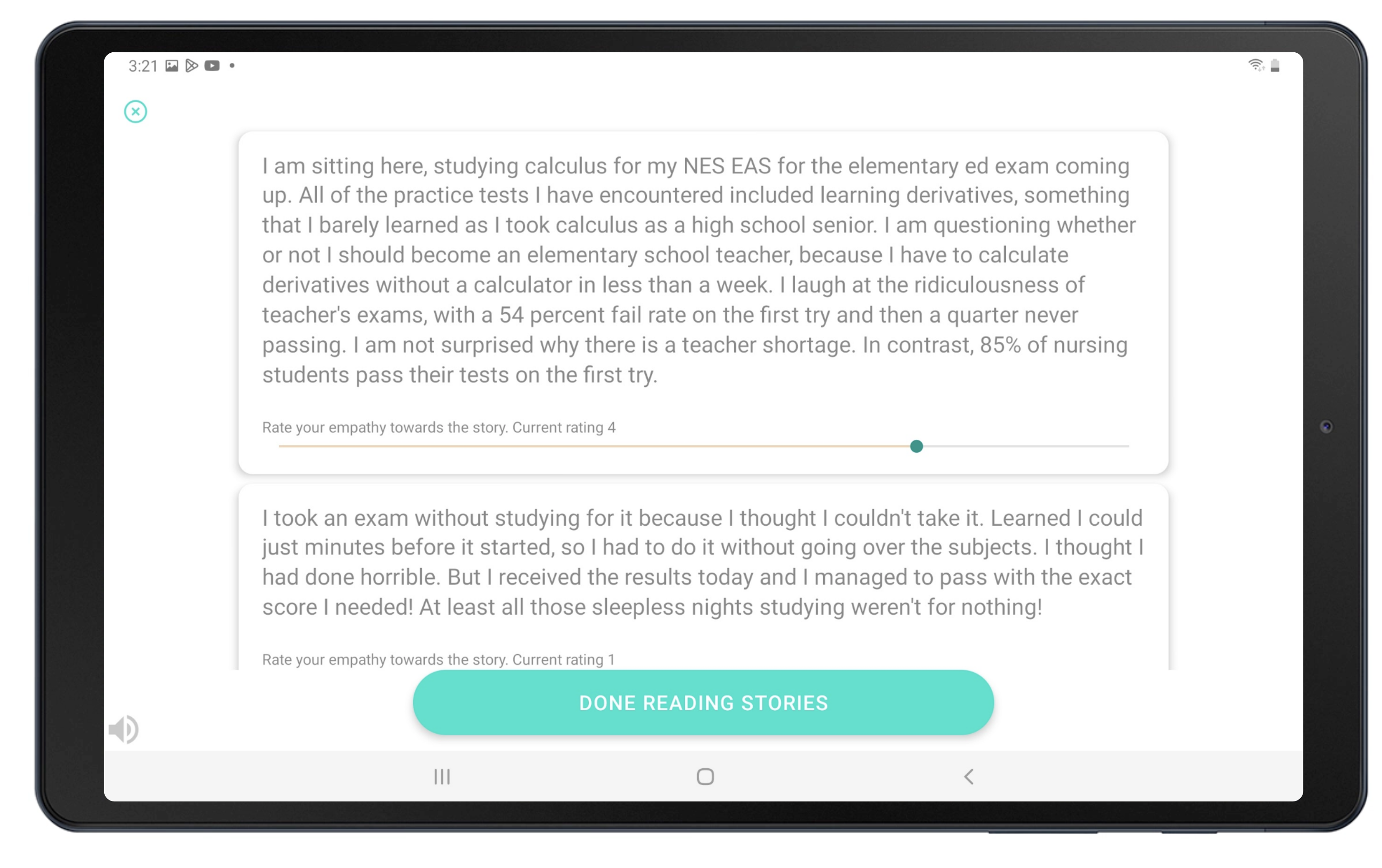} }}%
    \caption{(a) System overview of the robot station, which includes a tablet, Intel NUC computer, USB microphone, web camera, and Jibo social robot mounted together via a custom stand. (b) Stories are presented to the user on the tablet feed. }%
    \label{fig:example}%
    \vspace{-5pt}
\end{figure*}

\begin{enumerate}[left=0pt]
    \item \textbf{Warm up phase.} At the beginning of each section, the robot initiates casual conversation about the user's day or previous interactions. 

    \item \textbf{Story share phase.} The robot prompts the user to share a meaningful story based on what they wrote in their journal or if they had any other topics in mind.

    \item \textbf{Story receive phase.} The robot addresses the user's shared story by responding empathetically, then retrieves 3 stories the user might empathize with. It first summarizes the 3 stories verbally before directing the participant to read the full stories on the tablet. Whether the robot shares the stories as its own, e.g., ``I once ...", or as a social proxy, e.g., ``I know a story about a person who ...", depends on the study condition. On the tablet, the user rates their empathy toward each story using a slider on a scale of 1-5. Note that the story presented on the tablet supplemented the robot storytelling out loud -- the robot \textit{always} shared summaries of the stories verbally in addition to the tablet display.

    \item \textbf{Story reflection phase.} We carefully designed reflection prompts based on narrative therapy approaches, allowing participants to externalize their experiences through the story they read, in addition to relating to the narrator of the story \cite{gardner_one_2009, white_narrative_1990, yoosefi_looyeh_treating_2014}. The robot asks the participant to reflect on the following four areas: ways in which they related to the narrator, identifying the emotions of the narrator, regulating or comforting the narrator, and high-level takeaways from the story that the participant could apply to their own life.

    \item \textbf{Cool-down phase.} Finally, the agent summarizes the session, thanks the participant for their openness, and expresses excitement to chat again.
\end{enumerate}




\section{System Implementation} \label{systemimplementation}

\subsection{Social Robot Platform}
Our platform consists of the social robot (Jibo), an Android tablet as a shared workspace to display stories and questionnaires, and an Ubuntu machine with an external USB microphone and webcam (Fig.~\ref{fig:example}). The Firebase Realtime Database is used to pass sensor and software results between all hardware components and to store session metadata.
Sensitive information and data such as videos, audios, and transcriptions are synced to our institution's secure server. Participants were free to use the robot's out-of-the-box capabilities, which include playing games and chit-chat conversation, allowing the robot to serve as an always-on companion. The Jibo platform includes built in robot gesture, vocal articulation, and emojis displayed on the robot's face during story sharing, affording a multi-modal and embodied conversational experience. 

We developed the ``Share Your Story" app for participants to share  and view stories. As users speak to the robot, the audio recordings from the microphone are sent to AssemblyAI's real-time transcription service.
After the user signals the end of their turn through the tablet, these transcriptions are inputted into ChatGPT to generate a unique, personalized response.
These responses are relayed to the user through the robot's embodied speech abilities that synchronize its text-to-speech with appropriate body motion and facial expressions. We detail the conversational prompt implementation in Appendix \ref{prompts}. 




\subsection{Story Retrieval Model and Story Database} 

The primary focus of our interaction is on story sharing facilitated through the robot. As such, it was crucial to ensure the robot (1) maintains a database of diverse and high-quality personal narratives, and (2) retrieves relevant stories for the participant based on the conversation context. For our database of high-quality personal narratives, we use the \textsc{EmpathicStories} dataset \cite{shen_modeling_2023}, which contains 1,568 personal stories scraped from social media, crowdsourcing, and transcribed podcasts. Appendix \ref{fig:topics} shows clusters of overall story topics the robot can share, ranging from personal mental health struggles to life and career transitions. To retrieve emotionally resonant stories between the user's story and other people's personal stories, we adopted a BART-base model fine-tuned with the \textsc{EmpathicStories} dataset~\cite{shen_modeling_2023}. We adopted this model because the model outperforms ChatGPT and off-the-shelf models such as Sentence-BERT \cite{reimers_sentence-bert_2019} in retrieving empathetically relevant stories. 
We designed our robot to use this model to retrieve stories the user could relate to from the corpus of \textsc{EmpathicStories} when given a user's own story.

\section{Experiment}

\subsection{Participants and Study Design}

In total, $N=46$ participants were recruited as part of our study (Table \ref{demographics}). Of the 46 participants, 40 filled out the post-study survey and conducted the post-study interview (after 2 weeks). Six participants dropped out due to life changes such as changing jobs or moving cities (4), new time commitments from starting school (1), and privacy concerns (1). 

\begin{table}[h]
    \centering
    \footnotesize
    \caption{Participants' demographic breakdown (mean, s.d., range). No significant differences exist across conditions, \highlight{as verified by Mann-Whitney U tests}.}
    \resizebox{\linewidth}{!}{%
    \begin{tabular}{c|c|c}
\hline \textbf{Variable} & \textbf{Social Proxy (N = 19)} & \textbf{Social Agent (N = 21)} \\
\hline \textbf{Gender} & $\begin{array}{l}\text { Female 11}, 
\text { Male 8}\end{array}$ & $\begin{array}{l}\text { Female 13}, 
\text { Male 8}\end{array}$ \\
\hline \textbf{Age} & $\begin{array}{l}35.94  \pm 16.49,  [20.0, 75.0]\end{array}$ & $\begin{array}{l}34.38  \pm 12.78, [19.0, 68.0]\end{array}$ \\
\hline \textbf{Absorption} & $\begin{array}{l}3.64  \pm 0.88, [2, 5] \end{array}$ & $\begin{array}{l}3.62  \pm 0.77, [1.75, 4.75] \end{array}$ \\
\hline \textbf{Trait Empathy} & $\begin{array}{l}4.11  \pm 1.10,  [1, 5] \end{array}$ & $\begin{array}{l}4.38  \pm 0.80, [2, 5]\end{array}$ \\
\hline \textbf{Personality (Neuroticism)} & $\begin{array}{l}2.92  \pm 0.67, [2, 5] \end{array}$ & $\begin{array}{l}2.74  \pm 0.79, [1, 3.88]\end{array}$ \\
\hline
\end{tabular}}
\vspace{-5pt}
    \label{demographics}
\end{table}

We conducted a between-subject study where participants are subject to one of two conditions over the course of 2 weeks. In the \textbf{social agent} condition, after the participant shares their own personal story, the robot retrieves an emotionally relevant story and presents the story as its own. In the \textbf{social proxy} condition, the robot retrieves a personal story and presents it as another human narrator's story. \highlight{While many possible conditions could be tested (for example comparing to social connection users feel on social media, or to a disembodied condition such as journaling/reading suggested stories), the goal of our work is not to compare all possible configurations of social robots that mediate social connection, but to understand our new design of social proxy agents in-the-wild. As such, we chose the \textbf{social agent} condition where the robot self-discloses its own experiences, which is a common existing paradigm by which social robots aim to elicit empathy from users \cite{spitale_socially_2022, tsumura_influence_2023}. Many prior works have run controlled studies to explore the effects of [\textbf{D1}] - [\textbf{D4}], albeit not in social proxy settings, but few have specifically analyzed the effect of clarifying story ownership for narrative authenticity [\textbf{D5}], which is the main thrust of our social proxy paradigm. 
Through our deployment study, we use qualitative insights to understand user's experiences with the social proxy design.}

Across conditions, how the robot referred to and reflected on the story throughout the \textit{entire session} was different based on whether the robot was directed to be its own social agent or serve as a social proxy. We randomly assigned participants to a condition, then ran statistical tests to balance participants across conditions by demographic factors.



\subsection{Study Procedure}

We recruited participants over the age of 18 through online advertisements distributed through social media and mailing lists. Our study protocol was approved by our university's Institutional Review Board, and all participants filled out a consent form prior to starting the study, upon which they were asked to fill pre-study questionnaires. Then, participants were shipped a robot station (including a Jibo robot and Android tablet mounted together), and we scheduled 30-minute onboarding calls to set up the participant's station and explain the study. We also provided a user manual detailing robot station troubleshooting and the research team's contact information in case any technical difficulties might arise. 

Participants were asked to engage with the robot in the story sharing activity for at least 6 sessions over the course of two weeks (3 sessions per week). Note that we did not restrict the number of sessions so that future work can explore usage effects on robot engagement/perception, but find that the number of sessions is not significantly different across conditions ($p = 0.44$). 
 Outside the sessions, we encouraged participants to journal about any stories they would want to share with the robot during the ``Share Your Story" activity. We provided participants with a notebook to journal in, as well as a card of journal prompts based on narrative therapy approaches \cite{yoosefi_looyeh_treating_2014} (Appendix \ref{materials}). We administered surveys at the end of the 2 weeks and finally conducted 30-minute interviews remotely. Participants were compensated \$30 for 6 sessions.


\subsection{Data Collection and Analysis}

\subsubsection{Interviews}
We transcribed all interview recordings for qualitative analysis, where we thematically analyzed interviews using a reflexive, bottom-up approach via the open-coding method \cite{braun_using_2006} \highlight{in order to avoid preemptively prescribing our design insights to our qualitative analysis, and allowing the data to speak towards the design guidelines without researcher bias. }Two interview raters reviewed each category and formed sub-categories based on recurring themes present in the utterances. The interview raters then independently applied the coding scheme to the utterances (we achieve a substantial Cohen's Kappa inter-rater agreement of 0.74). A second round of coding and discussion was conducted on disagreements in order to determine the final codes. Ultimately, we constructed 14 overarching codes and condensed these codes into 3 themes. Themes and descriptions are presented in Table \ref{tab:themes}, and the full codebook is included in the Appendix.

\subsubsection{Self-Report Measures}
During our pre-study questionnaires, we administered the Big 5 Personality Test \cite{goldberg_structure_1993}, absorption scale dimensions of the Multidimensional Personality Questionnaire  
\cite{noauthor_multidimensional_nodate}, the Single Item Trait Empathy Scale \cite{konrath_development_2018}, Compassionate Love for Humanity Scale \cite{sprecher_compassionate_2005}, and UBC State Social Connection Scale \cite{lok_ubc_2022}. Note that we use both the Compassionate Love for Humanity Scale and the UBC State Social Connection to measure overall ``social connectedness.'' These surveys focus on different aspects of human connection, with Compassionate Love for Humanity capturing compassion towards \textit{strangers}, whereas the UBC State Social Connection scale captures social connection to both strangers and social partners. Within each session, participants rated how much they empathized with a story they read on a scale of 1-5.
At the end of the study, we additionally administered the Working Alliance Inventory (WAI) \cite{munder_working_2010}, the dyadic trust scale \cite{larzelere_dyadic_1980}, the godspeed surveys \cite{bartneck_measurement_2009}, robot perceived empathy, and surveys from \cite{spitale_socially_2022} to see if any changes in robot perception played a role in the effects in connection. 

\subsubsection{Video and Audio Recordings}
During the study, each station recorded the entire interaction session using the station's built in webcam and condenser microphone for high-quality recordings of the participant's face and voice. Note that we made clear when the system was recording the user in our study onboarding and through the robot's ring light. \highlight{Participants completed on average 7 sessions throughout the course of the study (7.17 ± 1.46, range 6-13), with an average session length of 11.33 minutes (11.33 ± 4.51, range 4.46-28.41), and an average of 3637.92 words per session (3637.92 ± 2579.5, range 450-15035). Further details of the dataset are published in \cite{shen-etal-2024-empathicstories}.}

\begin{table}[t]
\centering
\footnotesize
\caption{Description of qualitative analysis themes that emerged from open-coding.}
\label{tab:themes}
\resizebox{\linewidth}{!}{ 

\begin{tabularx}{\linewidth}{p{0.2\linewidth}|p{0.4\linewidth}|p{0.2\linewidth}}  
\hline
\textbf{Theme} & \textbf{Theme Description} & \textbf{Codes} \\
\hline
Social Connection and Empathy & People mention feelings of social connection and empathy with others while interacting with the robot, such as through drawing connections to stories or serving as a catalyst for other connections. & identifying connections, connection beyond stories, empathy and perspective taking, diverse perspectives \\
\hline
Positive robot qualities enhancing the interaction & The participants perceived specific positive qualities of the robot during interaction, such as the robot being non-judgmental or helping to reframe thoughts. & judgment-free, reframing, active listening, counseling, encouraging \\
\hline
Improvements around robot story sharing or form & The participants disliked qualities of either the way the robot shared stories or the form factor of the robot. & question authenticity, disagreement, repetition, mechanical, security \\
\hline
\end{tabularx}
}
\vspace{-10pt}
\end{table}


\section{Results}
\highlight{In the following section, we present qualitative insights from user interviews, organized by themes that emerged from bottom-up coding. In Section \ref{designguidelines}, we map our findings and codes to the specific design principles [\textbf{D1}] - [\textbf{D5}].}

\subsection{Main Effects on Social Connection and Empathy}

We first focus on results emerging from the ``Social Connection and Empathy'' theme (Table \ref{tab:themes}).
Participants commented on specific ways the robot system helped them connect with others, whether through identifying connections to other stories, fostering connections beyond stories, helping to put participants in other people's shoes, or offering diverse perspectives. Quotes below are shared from participants in the \textbf{social proxy} condition. 

Firstly, several participants were surprised by the robot's ability to draw connections between their own life and another person's life, for example, P39  mentioned,   \textit{``[the robot] was able to connect my story to others, which [was] really cool to see. It was almost  human like I was able to, empathize with my story and see the connections between them."} Another participant, P38, shared a specific memory of the robot making abstract connections to other people's stories, such as drawing resonance between their story about math to another person's story about architecture: ``\textit{College came up and [the robot] asked me what I'm studying, and I said math, and [he] asked me why I like math. And I was just like, you build up theory to solve things and sometimes you're creating all these abstractions...[the robot] somehow said, that, that sounds a lot like architecture. I can bring out the connections between math and architecture more. And that was the first time I realized [the robot], is really smart.}''

Secondly, other participants felt that the robot motivated them to socialize and connect beyond the robot interaction setting, either serving as a conversation topic for common ground, or sharing stories that resonate and reminded participants to reach out to others. For example, P52 mentioned, ``\textit{...with [my brother] we experiment with the robot ...I've created a bond between me and my brother, my brother and the robot, and the robot and me, and then that became more of like a conversational point.}" The robot's shared stories also served as resonant moments to remind participants of connections in their own life. P33 cited: ``\textit{Actually there was a time that helped me reconnect with a friend recently ... the questions [the robot] asked made me realize these stories are a lot deeper ... that influenced me to reach out to a friend.}''

Finally, participants expressed the ability to empathize and connect with people better after interacting with the robot by placing themselves in the shoes of others or being exposed to diverse perspectives. For example, P18 mentioned ``\textit{The robot asked questions like, what would you do if you were in this person's situation or how would you feel if you were in this person's situation? That kind of question really made me [imagine] me in that situation...I feel like I could relate to other people more easily, and I'm being more empathetic person in general.}''



Participants in the \textbf{social agent} condition shared that the robot interaction boosted their mood to help them approach in-person social interactions. For example, P03 shared, ``\textit{I have a good energy leaving the session and then go out with people or go network and socialize with other colleagues.}” Other participants expressed that the interaction was generally helpful for hearing about different stories and being exposed to different perspectives. For example, P22 stated: ``\textit{Something I miss a lot about college and just like being in a bigger city is like you meet people, you learn about so many different stories, different lives, um, and you're excited...So talking to the robot helps boost that area of social interaction.}'' Another participant (P47) similarly expressed ``\textit{it's interesting to get a different perspective and  a different person's experience, and then that helps get a different view of what you're struggling with or what your story was}.'' Through ``absorbing'' into the stories or ``stepping into other people's shoes," other participants found the robot's shared stories as reminders to empathize with others more broadly. For example, P20 shared, ``\textit{I think those moments of perseverance in the story...I definitely think that there were some great reminders of how to be in community with somebody or empathetic or supportive.}''

\highlight{While our study mainly aims to qualitatively understand the impact of social proxy agent design on social connection and empathy towards other people's stories, Appendix \ref{supplementarysurveyanalysis} includes supplementary results directly comparing survey measures of connection across the \textbf{social proxy} and \textbf{social agent} conditions via quantitative approaches. In summary, we found that participants showed more significant increases in connection and empathy towards stories when the robot delivered stories as a social proxy, rather than delivering as its own experiences.  }



\subsection{Effects of Robot Perception on Connection and Empathy}

In the results presented below, we focus on the themes of ``Positive robot qualities enhancing the interaction'' and ``Improvements around robot story sharing form'' (Table \ref{tab:themes}) in order to better understand how design and perception of the robot impacted social connection and empathy.

\paragraph{Robot Positive Qualities that Led to Better Interaction Experiences}
Participants overall found that the robot served as a judgment-free conversational partner, was effective in reframing their thoughts as well as providing active-listening, counseling, and encouragement. Firstly, participants found that the robot provided a ``judgment-free zone'' (P27) and was ``\textit{very good at ... open ended, non-judgmental, non directive questions}'' (P50), which made it helpful to articulate thoughts as well as personal stories. For example, P03 expressed that the robot helped foster more authentic self disclosure: ``\textit{I'm less anxious about talking to a robot because there's fear of judgment when I talk to a person}.'' A few participants expressed that the robot accommodated their personality, creating a channel for deep memories that they had not thought about before. For example, P15 said, the system was inclusive, and``\textit{It's a great idea for...providing it [to] people who don't like to talk to a person...because I'm really very shy and it's hard for me to start talking to someone you know...it was cathartic for me to get some of these old memories out especially of my brother because he's no longer here.}''

The participants also appreciated the robot’s ability to synthesize and recall information from sessions, which made participants feel like the robot really listened to them. For example, P36 shared ``\textit{... the robot understands my responses entirely and a lot of the conversation we have is reminiscent of talk therapy I have experienced in counseling sessions.}'' This effect was further mentioned by other participants who found the robot to provide helpful cognitive reframing to support people through emotional situations. For example, P24 ``\textit{I caught myself really thinking about his questions, realizing that he brought me to a place I had not brought myself.}''

Finally, participants mentioned that the robot's personality and form enhanced feelings of support and encouragement, which made them feel better after the interaction. 
For example, ``\textit{the gestures of the the face shape and head tilt indicated sort of agreement, but also more encouragement. I found these gestures very effective}'' (P52). Other participants mentioned feeling support through the robot's presence of ``always being there'' and navigating support in an adaptive manner. For example, 
P29 shared ``\textit{[the robot] always makes me feel better because [it] can walk a very fine line between unconditional love and support and ``hey well maybe here's  my suggestion for you,'' and I thought that was extremely impressive}.'' 

\highlight{Interestingly, our qualitative insights around positive qualities that enhanced interaction with the robot are supported by quantitative analysis -- We observed that the robot's perceived empathy is significantly positively correlated with changes in social connection from our survey measures ($\rho = 0.34$). }



\paragraph{Improvements About Robot Sharing and Form}


With regards to dislikes about the interactions, participants primarily commented on questioning the consistent persona of stories shared by the robot and disagreeing with the ways the story relates to their experiences. For example, P43 stated: ``\textit{I kind of lose trust in the authenticity of what [the robot] says...I feel like the stories are from many people scattered around, rather than having a conscious. It does not feel believable, and that's why if all the stories felt like they were from the same entity, I would maybe morph and make it more believable to me.}''

Participants additionally mentioned that social errors made by the robot, such as over-repetition of questions or stories, as well as unnatural speech or gestures, reduced the quality of the interaction. For example, P27  mentioned, ``\textit{The hardest part for me to get past on an empathy level is that his voice doesn't necessarily sound natural.}''

Finally, a few participants stated that they had to adjust to hosting the robot, as the robot would sometimes randomly start talking, which would be startling. For example, P13  said,  ``\textit{I guess the cameras turn on every six hours...it's not something that I'm actively worried about, but I do feel like that's probably the only discomfort.}''

\section{Discussion}

\subsection{The Impact of Social Proxy Robots}

Overall, participant interviews reveal that the social proxy interaction enhanced empathy and social connection via different mechanisms such as connecting personal experiences to other people's lives or encouraging social support beyond the interaction (\textbf{RQ1}). 
Participants commented that feelings of connection were enhanced through the robot via pathways such as drawing connections to other stories or offering diverse but relevant perspectives. 
Prior works indicate that framing the robot as a machine rather than a social agent and being clear about a robot's lack of human psychological capabilities reduces social presence and weakens the participants' relationship to the robot \cite{straten_transparency_2020, noauthor_effects_nodate}. Compared to these prior works, we focus on how this framing  affects relationship to \textit{other} people, \highlight{and show that the social proxy design provides new ways by which users are able to connect with others' experiences. }

\subsection{Perception of the Robot Influences Social Connection}
While prior works assess how stories can be delivered through the text modality alone to facilitate connection and behavioral changes \cite{bhattacharjee_i_2022}, social robots offer novel design affordances that can play a role in users' ability to connect with stories shared by the robot (\textbf{RQ2}). 
Our qualitative findings reveal that 
participants cite the robot's ``judgment-free'' nature, or being an ``objective third-party'' as factors that helped them more authentically share and take suggestions from the robot. The embodied form of the agent encouraged participants to step into the shoes of the narrator of a story and engage more deeply with the narrative.
On the other hand, participants mention that the robot's mechanical nature sometimes hindered their ability to fully empathize with stories, suggesting that more work on physical expression of stories is needed to fully maximize the embodied form of the social robot.

\section{Design Guidelines for Social Robots as Social Proxies}\label{designguidelines}

We draw design recommendations for future social proxy agents that mediate empathetic exchanges, revisiting the design principles presented in Section \ref{designprinciples} based on our findings, \highlight{and mapping principles to relevant codes that emerged in our qualitative analysis. }

\begin{itemize}[left=0pt]
    \item \highlight{\textbf{[D1] Social proxies with multi-modal interaction and physical embodiment could support more engaging interaction and greater emotional involvement in connection} [\textit{encouraging, mechanical, security}] --  Empathetic interactions necessitate expressive speech. While many robot text-to-speech designs are intentionally robotic to avoid the uncanny valley \cite{romportl_speech_2014}, participants mentioned in interviews that the robot's mechanical voice hindered empathy towards the stories. Furthermore, supporting physical embodiment necessitates more multimodal signals, which can cause security concerns raised by some participants. However, we found that when the robot successfully demonstrated social intelligence (such as nodding), the participants did feel more acknowledged and supported.
    Future social robots should incorporate sophisticated speech synthesis technologies with dynamic voice modulation to better convey emotional subtleties in the stories \cite{triantafyllopoulos_overview_2023}. Physical gestures and gaze could also be adapted in real-time to emphasize the tonality of the story \cite{huang_emotion_2024}, but should tradeoff with user privacy \cite{tonkin_embodiment_2017}. These elements are critical for achieving narrative transportation, where users feel fully absorbed in the world of the story, fostering greater empathy and connection \cite{johnson_transportation_2012}.}
    \item \highlight{\textbf{[D2] The non-judgmental nature of social proxy increased people's willingness to self-disclose.} [\textit{judgement-free}] -- In line with prior works \cite{lucas_its_2014, bethel_using_2016-1}, we observed that participants feel more comfortable disclosing personal information with the robot, quoting the robot's non-judgmental nature. Participants commented that this allowed them to share stories more authentically, and even share information that they ``would not otherwise share with another person.'' 
    Future social proxy interactions that aim to enhance connection should consider this design principle, and understand further how to evoke more authentic story sharing on social platforms, for example by explicitly emphasizing robot's neutrality, empathetic conversation capabilities, and transparency about how the message is used \cite{laban_opening_2023}.}

    \item \highlight{\textbf{[D3] Long-term relational interaction  help users feel heard.} [\textit{active listening, repetition}]-- In line with previous literature, we found that long term interaction is important for building a relationship with the agent \cite{kory-westlund_long-term_2022}, and \textit{memory} is foundational to supporting empathy and social connection. While active listening is often a real-time interaction experienced through back-channeling \cite{park_telling_2017}, we found in our study that participants ``felt heard'' when the robot remembered previous conversations and synthesized information across stories. However, overly repetitive dialogue across sessions made the interaction feel more canned, which could reduce user engagement. As such, future social proxy robots should include memory systems to track user inputs and use them to personalize future interactions \cite{paplu_harnessing_2022}. For instance, recalling a user’s shared story and connecting it to a newly retrieved narrative can reinforce a sense of continuity and foster deeper bonds.}
    \item \highlight{\textbf{[D4] Adaptive and socially contextualized conversation across stories and reflections.} [\textit{reframing, counseling, empathy and perspective taking, connection beyond stories, disagreement}]-- 
    Participants commented that the robot provided relevant counseling, despite counseling not being a direct design goal of our work. We found that flexible questions and stories shared by the robot aided in human connection forming by scaffolding the empathizing process, such as explicitly asking the user to put themselves in the story owner's shoes. Participants noted that relevant questioning from the robot actually influenced them to connect with friends or family in the real world. However, when the user disagreed with the story the robot chose, this weakened empathy. Future designs can draw on adaptive methods from the field of social language processing for improved empathic story matching and conversation generation towards goals such as wellbeing, connection, and flourishing\cite{zhang_towards_2024, rashkin_towards_2019, sharma_computational_2020}.}

    
    \item \highlight{\textbf{[D5] Transparency in robot's role and story ownership foster connection.} [\textit{identifying connections, diverse perspectives, question authenticity}] -- Finally, aligning robot's role and story delivery style matter in social outcomes, as people empathize and show more connection when the robot makes clear its role as social proxy and that the stories come from a person, allowing users to draw connections to different \textit{human} perspectives. 
    When the robot delivered life stories from others in first-person perspective, the stories felt less believable or authentic. As such, social proxy robots should clearly differentiate their role as facilitators of human stories to avoid conflating the robot's persona with the story’s origin to maintain user trust and focus on empathy between the user and other human narrators \cite{spitale_socially_2022, giorgi_2023, shen_empathy_2024}.}
    
\end{itemize}


\section{Conclusion and Future Directions}
This work investigates how social robots can be used as social platforms to asynchronously facilitate the sharing of personal stories across people in order to improve empathy and connection. 
\highlight{From our real-world deployment study and interview analyses, we found that social proxy agents, which share stories across people, increased connection and empathy towards others via pathways such as identifying connections to related stories and facilitating connections in real life beyond the stories. We showed that the form and personality of the robot, such as being encouraging or serving as a non-judging agent further solidified the connective experiences of the social proxy robot interaction. }
Finally, we discussed how our empirical findings guide future designs of social robots used as social proxies for enhancing human connection.
Our work sheds light on exploring how social robots can serve \textit{new roles} that can improve connection and foster empathy \highlight{between people's lived experiences. Future works may aim to understand how our design insights should be understood across different user's traits as well as their personal experiences. For example, social proxy agents may be more effective when stories retrieved are more closely tied to the user's life experiences, or the interaction may be more effective for populations who have increased risk of loneliness, such as older adults. We encourage future exploration social proxy agent designs and supporting interactions that are inclusive and enriching for all people.}

\bibliographystyle{IEEEtran}  
\balance
\bibliography{sample-base}   

\begin{thebibliography}{10}
\providecommand{\url}[1]{#1}
\csname url@samestyle\endcsname
\providecommand{\newblock}{\relax}
\providecommand{\bibinfo}[2]{#2}
\providecommand{\BIBentrySTDinterwordspacing}{\spaceskip=0pt\relax}
\providecommand{\BIBentryALTinterwordstretchfactor}{4}
\providecommand{\BIBentryALTinterwordspacing}{\spaceskip=\fontdimen2\font plus
\BIBentryALTinterwordstretchfactor\fontdimen3\font minus \fontdimen4\font\relax}
\providecommand{\BIBforeignlanguage}[2]{{%
\expandafter\ifx\csname l@#1\endcsname\relax
\typeout{** WARNING: IEEEtran.bst: No hyphenation pattern has been}%
\typeout{** loaded for the language `#1'. Using the pattern for}%
\typeout{** the default language instead.}%
\else
\language=\csname l@#1\endcsname
\fi
#2}}
\providecommand{\BIBdecl}{\relax}
\BIBdecl

\bibitem{oviatt_technology_2021}
S.~Oviatt, ``\BIBforeignlanguage{en}{Technology as {Infrastructure} for {Dehumanization}: {Three} {Hundred} {Million} {People} with the {Same} {Face}},'' in \emph{\BIBforeignlanguage{en}{Proceedings of the 2021 {International} {Conference} on {Multimodal} {Interaction}}}.\hskip 1em plus 0.5em minus 0.4em\relax New York, NY, USA: Association for Computing Machinery, Oct. 2021, pp. 278--287.

\bibitem{buecker_is_2021}
S.~Buecker, M.~Mund, S.~Chwastek, M.~Sostmann, and M.~Luhmann, ``Is loneliness in emerging adults increasing over time? {A} preregistered cross-temporal meta-analysis and systematic review,'' \emph{Psychological Bulletin}, vol. 147, pp. 787--805, 2021.

\bibitem{jang_design_2023}
S.~Jang, K.-R. Lee, G.~Goh, D.~Kim, G.~Yun, N.~Kim, B.~Kim~Lux, C.-W. Woo, H.~Kim, and Y.-W. Park, ``Design and field trial of {EmotionFrame}: exploring self-journaling experiences in homes for archiving personal feelings about daily events,'' \emph{Human–Computer Interaction}, vol.~0, no.~0, pp. 1--26, Jun. 2023.

\bibitem{morelli_emerging_2015}
S.~A. Morelli, M.~D. Lieberman, and J.~Zaki, ``\BIBforeignlanguage{en}{The {Emerging} {Study} of {Positive} {Empathy}},'' \emph{\BIBforeignlanguage{en}{Social and Personality Psychology Compass}}, vol.~9, no.~2, pp. 57--68, 2015.

\bibitem{seppala_social_2013}
E.~Seppala, T.~Rossomando, and J.~R. Doty, ``\BIBforeignlanguage{en}{Social {Connection} and {Compassion}: {Important} {Predictors} of {Health} and {Well}-{Being}},'' \emph{\BIBforeignlanguage{en}{Social Research: An International Quarterly}}, vol.~80, no.~2, pp. 411--430, Jun. 2013.

\bibitem{eyssel_loneliness_2013}
F.~Eyssel and N.~Reich, ``Loneliness makes the heart grow fonder (of robots) — {On} the effects of loneliness on psychological anthropomorphism,'' in \emph{2013 8th {ACM}/{IEEE} {International} {Conference} on {Human}-{Robot} {Interaction} ({HRI})}, Mar. 2013, pp. 121--122.

\bibitem{pirhonen_can_2020}
J.~Pirhonen, E.~Tiilikainen, S.~Pekkarinen, M.~Lemivaara, and H.~Melkas, ``\BIBforeignlanguage{en}{Can robots tackle late-life loneliness? {Scanning} of future opportunities and challenges in assisted living facilities},'' \emph{\BIBforeignlanguage{en}{Futures}}, vol. 124, p. 102640, Dec. 2020.

\bibitem{berridge_companion_2023}
C.~Berridge, Y.~Zhou, J.~M. Robillard, and J.~Kaye, ``\BIBforeignlanguage{en}{Companion robots to mitigate loneliness among older adults: {Perceptions} of benefit and possible deception},'' \emph{\BIBforeignlanguage{en}{Frontiers in Psychology}}, vol.~14, p. 1106633, Feb. 2023.

\bibitem{nunez_effect_2019}
E.~Nunez, M.~Hirokawa, M.~Perusquia-Hernandez, and K.~Suzuki, ``Effect on social connectedness and stress levels by using a huggable interface in remote communication,'' in \emph{2019 8th International Conference on Affective Computing and Intelligent Interaction ({ACII})}, Sep. 2019, pp. 1--7.

\bibitem{narain_promoting_2020}
J.~Narain, T.~Quach, M.~Davey, H.~W. Park, C.~Breazeal, and R.~Picard, ``Promoting wellbeing with sunny, a chatbot that facilitates positive messages within social groups,'' in \emph{Extended Abstracts of the 2020 {CHI} Conference on Human Factors in Computing Systems}, ser. {CHI} {EA} '20.\hskip 1em plus 0.5em minus 0.4em\relax New York, NY, USA: Association for Computing Machinery, Apr. 2020, pp. 1--8.

\bibitem{rahwan_intelligent_2020}
I.~Rahwan, J.~W. Crandall, and J.-F. Bonnefon, ``\BIBforeignlanguage{en}{Intelligent machines as social catalysts},'' \emph{\BIBforeignlanguage{en}{Proceedings of the National Academy of Sciences}}, vol. 117, no.~14, pp. 7555--7557, Apr. 2020.

\bibitem{alves-oliveira_robots_nodate}
P.~Alves-Oliveira, E.~A. Bj{\"o}rling, P.~Wiesmann, H.~Dwikat, S.~Bhatia, K.~Mihata, and M.~Cakmak, ``Robots for connection: A co-design study with adolescents,'' in \emph{2022 31st IEEE International Conference on Robot and Human Interactive Communication (RO-MAN)}.\hskip 1em plus 0.5em minus 0.4em\relax IEEE, 2022, pp. 578--583.

\bibitem{ahmed_robots_2022}
E.~Ahmed, ``\BIBforeignlanguage{en}{Robots as {Human} {Companions}: {A} {Review}},'' \emph{\BIBforeignlanguage{en}{PACIS 2022 Proceedings}}, p.~18, 2022.

\bibitem{anzabi_exploring_2023}
N.~Anzabi, A.~Etemad, and H.~Umemuro, ``Exploring the {Effects} of {Self}-{Disclosed} {Backstory} of {Social} {Robots} on {Development} of {Trust} in {Human}-{Robot} {Interaction},'' in \emph{Companion of the 2023 {ACM}/{IEEE} {International} {Conference} on {Human}-{Robot} {Interaction}}, ser. {HRI} '23.\hskip 1em plus 0.5em minus 0.4em\relax New York, NY, USA: Association for Computing Machinery, Mar. 2023, pp. 431--435.

\bibitem{rosenthal-vonderputten_experimental_2013}
A.~M. Rosenthal-von der Pütten, N.~C. Krämer, L.~Hoffmann, S.~Sobieraj, and S.~C. Eimler, ``\BIBforeignlanguage{en}{An {Experimental} {Study} on {Emotional} {Reactions} {Towards} a {Robot}},'' \emph{\BIBforeignlanguage{en}{International Journal of Social Robotics}}, vol.~5, no.~1, pp. 17--34, Jan. 2013.

\bibitem{spitale_socially_2022}
M.~Spitale, S.~Okamoto, M.~Gupta, H.~Xi, and M.~J. Matarić, ``\BIBforeignlanguage{en}{Socially {Assistive} {Robots} as {Storytellers} {That} {Elicit} {Empathy}},'' \emph{\BIBforeignlanguage{en}{ACM Transactions on Human-Robot Interaction}}, p. 3538409, May 2022.

\bibitem{hancock_ai-mediated_2020}
J.~T. Hancock, M.~Naaman, and K.~Levy, ``{AI}-mediated communication: Definition, research agenda, and ethical considerations,'' \emph{Journal of Computer-Mediated Communication}, vol.~25, no.~1, pp. 89--100, 2020.

\bibitem{leong_social_2023-1}
J.~Leong, Y.~Teng, X.~B. Liu, H.~Jun, S.~Kratz, Y.~J. Tham, A.~Monroy-Hernández, B.~A. Smith, and R.~Vaish, ``Social wormholes: Exploring preferences and opportunities for distributed and physically-grounded social connections,'' \emph{Proc. {ACM} Hum.-Comput. Interact.}, vol.~7, pp. 359:1--359:29, 2023.

\bibitem{herrera_building_2018}
F.~Herrera, J.~Bailenson, E.~Weisz, E.~Ogle, and J.~Zaki, ``Building long-term empathy: A large-scale comparison of traditional and virtual reality perspective-taking,'' \emph{PloS one}, vol.~13, no.~10, p. e0204494, 2018.

\bibitem{oh_virtually_2016}
S.~Y. Oh, J.~Bailenson, E.~Weisz, and J.~Zaki, ``Virtually old: Embodied perspective taking and the reduction of ageism under threat,'' \emph{Computers in Human Behavior}, vol.~60, pp. 398--410, 2016.

\bibitem{mulvaney_social_2024}
P.~Mulvaney, B.~Rooney, M.~A. Friehs, and J.~F. Leader, ``Social {VR} design features and experiential outcomes: narrative review and relationship map for dyadic agent conversations,'' \emph{Virtual Reality}, vol.~28, no.~1, p.~45, 2024.

\bibitem{poon_computer-mediated_2023}
A.~Poon, M.~Luebke, J.~Loughman, A.~Lee, L.~Guerrero, M.~Sterling, and N.~Dell, ``Computer-mediated sharing circles for intersectional peer support with home care workers,'' \emph{Proc. {ACM} Hum.-Comput. Interact.}, vol.~7, pp. 39:1--39:35, 2023.

\bibitem{kucera_bedtime_2021}
J.~Kučera, J.~Scott, S.~Lindley, and P.~Olivier, ``Bedtime window: A field study connecting bedrooms of long-distance couples using a slow photo-stream and shared real-time inking,'' in \emph{Proceedings of the 2021 {CHI} Conference on Human Factors in Computing Systems}, ser. {CHI} '21.\hskip 1em plus 0.5em minus 0.4em\relax Association for Computing Machinery, 2021, pp. 1--12.

\bibitem{davis_promoting_2016}
K.~Davis, E.~Owusu, J.~Hu, L.~Marcenaro, C.~Regazzoni, and L.~Feijs, ``Promoting social connectedness through human activity-based ambient displays,'' in \emph{Proceedings of the International Symposium on Interactive Technology and Ageing Populations}, ser. {ITAP} '16.\hskip 1em plus 0.5em minus 0.4em\relax Association for Computing Machinery, 2016, pp. 64--76.

\bibitem{liu_significant_2021}
F.~Liu, C.~Park, Y.~J. Tham, T.-Y. Tsai, L.~Dabbish, G.~Kaufman, and A.~Monroy-Hernández, ``Significant otter: Understanding the role of biosignals in communication,'' in \emph{Proceedings of the 2021 {CHI} Conference on Human Factors in Computing Systems}, ser. {CHI} '21.\hskip 1em plus 0.5em minus 0.4em\relax Association for Computing Machinery, 2021, pp. 1--15.

\bibitem{wei_mediated_2023}
Q.~Wei, M.~Li, and J.~Hu, ``Mediated social touch with mobile devices: A review of designs and evaluations,'' \emph{{IEEE} Transactions on Haptics}, pp. 1--20, 2023.

\bibitem{k_miller_meeting_2021}
M.~K.~Miller, M.~Johannes~Dechant, and R.~L.~Mandryk, ``Meeting you, seeing me: The role of social anxiety, visual feedback, and interface layout in a get-to-know-you task via video chat.'' in \emph{Proceedings of the 2021 {CHI} Conference on Human Factors in Computing Systems}, ser. {CHI} '21.\hskip 1em plus 0.5em minus 0.4em\relax Association for Computing Machinery, 2021, pp. 1--14.

\bibitem{olsson_technologies_2020}
T.~Olsson, P.~Jarusriboonchai, P.~Woźniak, S.~Paasovaara, K.~Väänänen, and A.~Lucero, ``Technologies for enhancing collocated social interaction: Review of design solutions and approaches,'' \emph{Computer Supported Cooperative Work ({CSCW})}, vol.~29, no.~1, pp. 29--83, 2020.

\bibitem{stepanova_strategies_2022-1}
E.~R. Stepanova, J.~Desnoyers-Stewart, K.~Höök, and B.~E. Riecke, ``Strategies for fostering a genuine feeling of connection in technologically mediated systems,'' in \emph{Proceedings of the 2022 {CHI} Conference on Human Factors in Computing Systems}, ser. {CHI} '22.\hskip 1em plus 0.5em minus 0.4em\relax Association for Computing Machinery, 2022, pp. 1--26.

\bibitem{jeong_fribo_2018-2}
K.~Jeong, J.~Sung, H.~S. Lee, A.~Kim, H.~Kim, C.~Park, Y.~Jeong, J.~Lee, and J.~Kim, ``Fribo: 13th {Annual} {ACM}/{IEEE} {International} {Conference} on {Human}-{Robot} {Interaction}, {HRI} 2018,'' \emph{HRI 2018 - Proceedings of the 2018 ACM/IEEE International Conference on Human-Robot Interaction}, pp. 114--122, Feb. 2018.

\bibitem{sebo_robots_2020}
S.~Sebo, B.~Stoll, B.~Scassellati, and M.~F. Jung, ``Robots in {Groups} and {Teams}: {A} {Literature} {Review},'' \emph{Proceedings of the ACM on Human-Computer Interaction}, vol.~4, no. CSCW2, pp. 176:1--176:36, Oct. 2020.

\bibitem{strohkorb_sebo_strategies_2020}
S.~Strohkorb~Sebo, L.~L. Dong, N.~Chang, and B.~Scassellati, ``\BIBforeignlanguage{en}{Strategies for the {Inclusion} of {Human} {Members} within {Human}-{Robot} {Teams}},'' in \emph{\BIBforeignlanguage{en}{Proceedings of the 2020 {ACM}/{IEEE} {International} {Conference} on {Human}-{Robot} {Interaction}}}.\hskip 1em plus 0.5em minus 0.4em\relax Cambridge United Kingdom: ACM, Mar. 2020, pp. 309--317.

\bibitem{weisswange_what_2023}
T.~H. Weisswange, H.~Javed, M.~Dietrich, T.~V. Pham, M.~T. Parreira, M.~Sack, and N.~Jamali, ``\BIBforeignlanguage{en}{What {Could} a {Social} {Mediator} {Robot} {Do}? {Lessons} from {Real}-{World} {Mediation} {Scenarios}},'' in \emph{\BIBforeignlanguage{en}{IEEE ICRA WORKSHOP TOWARDS A BALANCED CYBERPHYSICAL SOCIETY: A FOCUS ON GROUP SOCIAL DYNAMICS}}, Jun. 2023.

\bibitem{breazeal_effects_2005}
C.~Breazeal, C.~Kidd, A.~Thomaz, G.~Hoffman, and M.~Berlin, ``Effects of nonverbal communication on efficiency and robustness in human-robot teamwork,'' in \emph{2005 {IEEE}/{RSJ} {International} {Conference} on {Intelligent} {Robots} and {Systems}}, Aug. 2005, pp. 708--713.

\bibitem{wainer_role_2006}
J.~Wainer, D.~J. Feil-seifer, D.~A. Shell, and M.~J. Mataric, ``The role of physical embodiment in human-robot interaction,'' in \emph{{ROMAN} 2006 - {The} 15th {IEEE} {International} {Symposium} on {Robot} and {Human} {Interactive} {Communication}}, Sep. 2006, pp. 117--122.

\bibitem{deng_embodiment_2019}
E.~Deng, B.~Mutlu, and M.~J. Mataric, ``\BIBforeignlanguage{English}{Embodiment in {Socially} {Interactive} {Robots}},'' \emph{\BIBforeignlanguage{English}{Foundations and Trends® in Robotics}}, vol.~7, no.~4, pp. 251--356, Jan. 2019.

\bibitem{ventre-dominey_embodiment_2019}
J.~Ventre-Dominey, G.~Gibert, M.~Bosse-Platiere, A.~Farnè, P.~F. Dominey, and F.~Pavani, ``\BIBforeignlanguage{en}{Embodiment into a robot increases its acceptability},'' \emph{\BIBforeignlanguage{en}{Scientific Reports}}, vol.~9, no.~1, p. 10083, Jul. 2019.

\bibitem{bethel_using_2016-1}
C.~L. Bethel, Z.~Henkel, K.~Stives, D.~C. May, D.~K. Eakin, M.~Pilkinton, A.~Jones, and M.~Stubbs-Richardson, ``Using robots to interview children about bullying: Lessons learned from an exploratory study,'' in \emph{2016 25th IEEE International Symposium on Robot and Human Interactive Communication (RO-MAN)}.\hskip 1em plus 0.5em minus 0.4em\relax IEEE, 2011, pp. 712--717.

\bibitem{pickard_revealing_2016}
M.~Pickard, C.~Roster, and Y.~Chen, ``Revealing sensitive information in personal interviews: {Is} self-disclosure easier with humans or avatars and under what conditions?'' \emph{Computers in Human Behavior}, vol.~65, pp. 23--30, Dec. 2016.

\bibitem{lucas_its_2014}
G.~M. Lucas, J.~Gratch, A.~King, and L.-P. Morency, ``\BIBforeignlanguage{en}{It’s only a computer: {Virtual} humans increase willingness to disclose},'' \emph{\BIBforeignlanguage{en}{Computers in Human Behavior}}, vol.~37, pp. 94--100, Aug. 2014.

\bibitem{bethel_secret-sharing_2011}
C.~L. Bethel, M.~R. Stevenson, and B.~Scassellati, ``Secret-sharing: Interactions between a child, robot, and adult,'' in \emph{2011 {IEEE} International Conference on Systems, Man, and Cybernetics}, 2011, pp. 2489--2494.

\bibitem{keen_theory_2022}
S.~Keen, ``\BIBforeignlanguage{en}{A {Theory} of {Narrative} {Empathy}},'' \emph{\BIBforeignlanguage{en}{NARRATIVE}}, p.~31, 2022.

\bibitem{hodges_giving_2010}
S.~D. Hodges, K.~J. Kiel, A.~D.~I. Kramer, D.~Veach, and B.~R. Villanueva, ``\BIBforeignlanguage{en}{Giving {Birth} to {Empathy}: {The} {Effects} of {Similar} {Experience} on {Empathic} {Accuracy}, {Empathic} {Concern}, and {Perceived} {Empathy}},'' \emph{\BIBforeignlanguage{en}{Personality and Social Psychology Bulletin}}, vol.~36, no.~3, pp. 398--409, Mar. 2010.

\bibitem{jeong_robotic_2023}
S.~Jeong, L.~Aymerich-Franch, S.~Alghowinem, R.~W. Picard, C.~L. Breazeal, and H.~W. Park, ``\BIBforeignlanguage{en}{A {Robotic} {Companion} for {Psychological} {Well}-being: {A} {Long}-term {Investigation} of {Companionship} and {Therapeutic} {Alliance}},'' in \emph{\BIBforeignlanguage{en}{Proceedings of the 2023 {ACM}/{IEEE} {International} {Conference} on {Human}-{Robot} {Interaction}}}.\hskip 1em plus 0.5em minus 0.4em\relax Stockholm Sweden: ACM, Mar. 2023, pp. 485--494.

\bibitem{laban_sharing_2023}
G.~Laban and E.~S. Cross, ``\BIBforeignlanguage{en}{Sharing with {Robots}: {Why} do we do it and how does it make us feel?}'' \emph{\BIBforeignlanguage{en}{IEEE Transactions on Affective Computing}}, vol.~PP, pp. 1--18, 01 2024.

\bibitem{westlund_measuring_2018}
J.~M.~K. Westlund, H.~W. Park, R.~Williams, and C.~Breazeal, ``\BIBforeignlanguage{en}{Measuring young children's long-term relationships with social robots},'' in \emph{\BIBforeignlanguage{en}{Proceedings of the 17th {ACM} {Conference} on {Interaction} {Design} and {Children}}}.\hskip 1em plus 0.5em minus 0.4em\relax Trondheim Norway: ACM, Jun. 2018, pp. 207--218.

\bibitem{kory2019exploring}
J.~M. Kory-Westlund and C.~Breazeal, ``Exploring the effects of a social robot's speech entrainment and backstory on young children's emotion, rapport, relationship, and learning,'' \emph{Frontiers in Robotics and AI}, vol.~6, p.~54, 2019.

\bibitem{kwak_what_2013}
S.~S. Kwak, Y.~Kim, E.~Kim, C.~Shin, and K.~Cho, ``What makes people empathize with an emotional robot?: {The} impact of agency and physical embodiment on human empathy for a robot,'' in \emph{2013 {IEEE} {RO}-{MAN}}, Aug. 2013, pp. 180--185.

\bibitem{costa_emotional_2018}
S.~Costa, A.~Brunete, B.-C. Bae, and N.~Mavridis, ``Emotional {Storytelling} {Using} {Virtual} and {Robotic} {Agents},'' \emph{International Journal of Humanoid Robotics}, vol.~15, no.~03, p. 1850006, Jun. 2018.

\bibitem{johnson_transportation_2012}
D.~R. Johnson, ``Transportation into a story increases empathy, prosocial behavior, and perceptual bias toward fearful expressions,'' \emph{Personality and Individual Differences}, vol.~52, no.~2, pp. 150--155, Jan. 2012.

\bibitem{utz_function_2015}
S.~Utz, ``The function of self-disclosure on social network sites: {Not} only intimate, but also positive and entertaining self-disclosures increase the feeling of connection,'' \emph{Computers in Human Behavior}, vol.~45, pp. 1--10, Apr. 2015.

\bibitem{laban_opening_2023}
G.~Laban, A.~Kappas, V.~Morrison, and E.~S. Cross, ``Opening up to social robots: How emotions drive self-disclosure behavior,'' in \emph{2023 32nd {IEEE} International Conference on Robot and Human Interactive Communication ({RO}-{MAN})}, 2023, pp. 1697--1704.

\bibitem{noguchi_how_2023}
Y.~Noguchi, H.~Kamide, and F.~Tanaka, ``How should a social mediator robot convey messages about the self-disclosures of elderly people to recipients?'' \emph{International Journal of Social Robotics}, vol.~15, no.~7, pp. 1079--1099, 2023.

\bibitem{shen_modeling_2023}
J.~Shen, M.~Sap, P.~Colon-Hernandez, H.~W. Park, and C.~Breazeal, ``Modeling {Empathic} {Similarity} in {Personal} {Narratives},'' in \emph{Proceedings of the 2023 Conference on Empirical Methods in Natural Language Processing}, H.~Bouamor, J.~Pino, and K.~Bali, Eds., May 2023, pp. 6237--6252.

\bibitem{kory-westlund_long-term_2022}
J.~M. Kory-Westlund, H.~Won~Park, I.~Grover, and C.~Breazeal, ``Long-{Term} {Interaction} with {Relational} {SIAs},'' in \emph{The {Handbook} on {Socially} {Interactive} {Agents}: 20 years of {Research} on {Embodied} {Conversational} {Agents}, {Intelligent} {Virtual} {Agents}, and {Social} {Robotics} {Volume} 2: {Interactivity}, {Platforms}, {Application}}, 1st~ed.\hskip 1em plus 0.5em minus 0.4em\relax New York, NY, USA: Association for Computing Machinery, Nov. 2022, vol.~48, pp. 195--260.

\bibitem{leite_long-term_2012}
I.~Leite, G.~Castellano, A.~Pereira, C.~Martinho, and A.~Paiva, ``\BIBforeignlanguage{en}{Long-{Term} {Interactions} with {Empathic} {Robots}: {Evaluating} {Perceived} {Support} in {Children}},'' in \emph{\BIBforeignlanguage{en}{Social {Robotics}}}, S.~S. Ge, O.~Khatib, J.-J. Cabibihan, R.~Simmons, and M.-A. Williams, Eds.\hskip 1em plus 0.5em minus 0.4em\relax Berlin, Heidelberg: Springer, 2012, pp. 298--307.

\bibitem{kasap_building_2012}
Z.~Kasap and N.~Magnenat-Thalmann, ``Building long-term relationships with virtual and robotic characters: the role of remembering,'' \emph{The Visual Computer}, vol.~28, no.~1, pp. 87--97, 2012.

\bibitem{niederhoffer_linguistic_2002}
K.~G. Niederhoffer and J.~W. Pennebaker, ``\BIBforeignlanguage{en}{Linguistic {Style} {Matching} in {Social} {Interaction}},'' \emph{\BIBforeignlanguage{en}{Journal of Language and Social Psychology}}, vol.~21, no.~4, pp. 337--360, Dec. 2002.

\bibitem{gouldner_norm_1960}
A.~W. Gouldner, ``The norm of reciprocity: A preliminary statement,'' \emph{American Sociological Review}, vol.~25, no.~2, pp. 161--178, 1960.

\bibitem{emerson_social_1976}
R.~M. Emerson, ``Social exchange theory,'' \emph{Annual Review of Sociology}, vol.~2, pp. 335--362, 1976.

\bibitem{bhattacharjee_i_2022}
A.~Bhattacharjee, J.~J. Williams, K.~Chou, J.~Tomlinson, J.~Meyerhoff, A.~Mariakakis, and R.~Kornfield, ``"{I} {Kind} of {Bounce} off {It}":{Translating} {Mental} {Health} {Principles} into {Real} {Life} {Through} {Story}-{Based} {Text} {Messages},'' \emph{Proceedings of the ACM on Human-Computer Interaction}, vol.~6, no. CSCW2, pp. 398:1--398:31, Nov. 2022.

\bibitem{pillemer_remembering_1992}
D.~B. Pillemer, ``Remembering personal circumstances: A functional analysis,'' in \emph{Affect and accuracy in recall: Studies of "flashbulb" memories}, ser. Emory symposia in cognition, 4.\hskip 1em plus 0.5em minus 0.4em\relax Cambridge University Press, 1992, pp. 236--264.

\bibitem{dinakar_you_2012}
K.~Dinakar, B.~Jones, H.~Lieberman, R.~Picard, C.~Rose, M.~Thoman, and R.~Reichart, ``You too?! mixed-initiative {LDA} story matching to help teens in distress,'' \emph{Proceedings of the International {AAAI} Conference on Web and Social Media}, vol.~6, no.~1, pp. 74--81, 2012.

\bibitem{shen_empathy_2024}
J.~Shen, D.~{DiPaola}, S.~Ali, M.~Sap, H.~W. Park, and C.~Breazeal, ``Empathy toward artificial intelligence versus human experiences and the role of transparency in mental health and social support chatbot design: Comparative study,'' \emph{{JMIR} Mental Health}, vol.~11, no.~1, p. e62679, 2024.

\bibitem{giorgi_2023}
S.~Giorgi, D.~M. Markowitz, N.~Soni, V.~Varadarajan, S.~Mangalik, and H.~A. Schwartz, ``\BIBforeignlanguage{en}{``{I} {Slept} {Like} a {Baby}'': {Using} {Human} {Traits} {To} {Characterize} {Deceptive} {ChatGPT} and {Human} {Text}},'' \emph{\BIBforeignlanguage{en}{Proceedings of the IACT’23 Workshop}}, 2023.

\bibitem{tsumura_influence_2023}
T.~Tsumura and S.~Yamada, ``Influence of agent’s self-disclosure on human empathy,'' \emph{{PLOS} {ONE}}, vol.~18, no.~5, p. e0283955, 2023.

\bibitem{gardner_one_2009}
P.~J. Gardner and J.~M. Poole, ``\BIBforeignlanguage{en}{One {Story} at a {Time}: {Narrative} {Therapy}, {Older} {Adults}, and {Addictions}},'' \emph{\BIBforeignlanguage{en}{Journal of Applied Gerontology}}, vol.~28, no.~5, pp. 600--620, Oct. 2009.

\bibitem{white_narrative_1990}
M.~White and D.~Epston, \emph{\BIBforeignlanguage{English}{Narrative {Means} to {Therapeutic} {Ends}}}, 1st~ed.\hskip 1em plus 0.5em minus 0.4em\relax New York: W. W. Norton \& Company, May 1990.

\bibitem{yoosefi_looyeh_treating_2014}
M.~Yoosefi~Looyeh, K.~Kamali, A.~Ghasemi, and P.~Tonawanik, ``\BIBforeignlanguage{en}{Treating social phobia in children through group narrative therapy},'' \emph{\BIBforeignlanguage{en}{The Arts in Psychotherapy}}, vol.~41, no.~1, pp. 16--20, Feb. 2014.

\bibitem{reimers_sentence-bert_2019}
N.~Reimers and I.~Gurevych, ``Sentence-{BERT}: Sentence embeddings using siamese {BERT}-networks,'' in \emph{Proceedings of the 2019 Conference on Empirical Methods in Natural Language Processing and the 9th International Joint Conference on Natural Language Processing ({EMNLP}-{IJCNLP})}.\hskip 1em plus 0.5em minus 0.4em\relax Association for Computational Linguistics, 2019, pp. 3980--3990.

\bibitem{braun_using_2006}
V.~Braun and V.~Clarke, ``Using thematic analysis in psychology,'' \emph{Qualitative Research in Psychology}, vol.~3, no.~2, pp. 77--101, Jan. 2006.

\bibitem{goldberg_structure_1993}
L.~R. Goldberg, ``The structure of phenotypic personality traits,'' \emph{American Psychologist}, vol.~48, no.~1, pp. 26--34, 1993.

\bibitem{noauthor_multidimensional_nodate}
A.~Tellegen and N.~G. Waller, ``\BIBforeignlanguage{en-US}{Multidimensional {Personality} {Questionnaire} ({MPQ})},'' \emph{\BIBforeignlanguage{en-US}{Addiction Research Center}}, 2008.

\bibitem{konrath_development_2018}
S.~Konrath, B.~P. Meier, and B.~J. Bushman, ``\BIBforeignlanguage{en}{Development and validation of the single item trait empathy scale ({SITES})},'' \emph{\BIBforeignlanguage{en}{Journal of Research in Personality}}, vol.~73, pp. 111--122, Apr. 2018.

\bibitem{sprecher_compassionate_2005}
S.~Sprecher and B.~Fehr, ``\BIBforeignlanguage{en}{Compassionate love for close others and humanity},'' \emph{\BIBforeignlanguage{en}{Journal of Social and Personal Relationships}}, vol.~22, no.~5, pp. 629--651, Oct. 2005.

\bibitem{lok_ubc_2022}
I.~Lok and E.~Dunn, ``\BIBforeignlanguage{en}{The {UBC} {State} {Social} {Connection} {Scale}: {Factor} {Structure}, {Reliability}, and {Validity}},'' \emph{\BIBforeignlanguage{en}{Social Psychological and Personality Science}}, p. 19485506221132090, Nov. 2022.

\bibitem{munder_working_2010}
T.~Munder, F.~Wilmers, R.~Leonhart, H.~W. Linster, and J.~Barth, ``\BIBforeignlanguage{eng}{Working {Alliance} {Inventory}-{Short} {Revised} ({WAI}-{SR}): psychometric properties in outpatients and inpatients},'' \emph{\BIBforeignlanguage{eng}{Clinical Psychology \& Psychotherapy}}, vol.~17, no.~3, pp. 231--239, 2010.

\bibitem{larzelere_dyadic_1980}
R.~E. Larzelere and T.~L. Huston, ``The {Dyadic} {Trust} {Scale}: {Toward} {Understanding} {Interpersonal} {Trust} in {Close} {Relationships},'' \emph{Journal of Marriage and Family}, vol.~42, no.~3, pp. 595--604, 1980.

\bibitem{bartneck_measurement_2009}
C.~Bartneck, D.~Kulić, E.~Croft, and S.~Zoghbi, ``\BIBforeignlanguage{en}{Measurement {Instruments} for the {Anthropomorphism}, {Animacy}, {Likeability}, {Perceived} {Intelligence}, and {Perceived} {Safety} of {Robots}},'' \emph{\BIBforeignlanguage{en}{International Journal of Social Robotics}}, vol.~1, no.~1, pp. 71--81, Jan. 2009.

\bibitem{shen-etal-2024-empathicstories}
J.~Shen, Y.~Kim, M.~Hulse, W.~Zulfikar, S.~Alghowinem, C.~Breazeal, and H.~Park, ``{E}mpathic{S}tories++: A multimodal dataset for empathy towards personal experiences,'' in \emph{Findings of the Association for Computational Linguistics: ACL 2024}, L.-W. Ku, A.~Martins, and V.~Srikumar, Eds.\hskip 1em plus 0.5em minus 0.4em\relax Bangkok, Thailand: Association for Computational Linguistics, Aug. 2024, pp. 4525--4536.

\bibitem{straten_transparency_2020}
C.~L.~V. Straten, J.~Peter, R.~Kühne, and A.~Barco, ``\BIBforeignlanguage{en}{Transparency about a {Robot}'s {Lack} of {Human} {Psychological} {Capacities}: {Effects} on {Child}-{Robot} {Perception} and {Relationship} {Formation}},'' \emph{\BIBforeignlanguage{en}{ACM Transactions on Human-Robot Interaction}}, vol.~9, no.~2, pp. 1--22, Jun. 2020.

\bibitem{noauthor_effects_nodate}
J.~M. Kory~Westlund, M.~Martinez, M.~Archie, M.~Das, and C.~Breazeal, ``\BIBforeignlanguage{en-US}{Effects of framing a robot as a social agent or as a machine on children's social behavior},'' \emph{\BIBforeignlanguage{en-US}{IEEE International Workshop on Robot and Human Communication (ROMAN)}}, 2016.

\bibitem{romportl_speech_2014}
J.~Romportl, ``\BIBforeignlanguage{en}{Speech {Synthesis} and {Uncanny} {Valley}},'' in \emph{\BIBforeignlanguage{en}{Text, {Speech} and {Dialogue}}}, ser. Lecture {Notes} in {Computer} {Science}, P.~Sojka, A.~Horák, I.~Kopeček, and K.~Pala, Eds.\hskip 1em plus 0.5em minus 0.4em\relax Cham: Springer International Publishing, 2014, pp. 595--602.

\bibitem{triantafyllopoulos_overview_2023}
A.~Triantafyllopoulos, B.~W. Schuller, G.~İymen, M.~Sezgin, X.~He, Z.~Yang, P.~Tzirakis, S.~Liu, S.~Mertes, E.~André, R.~Fu, and J.~Tao, ``An overview of affective speech synthesis and conversion in the deep learning era,'' \emph{Proceedings of the {IEEE}}, vol. 111, no.~10, pp. 1355--1381, 2023.

\bibitem{huang_emotion_2024}
P.~Huang, Y.~Hu, N.~Nechyporenko, D.~Kim, W.~Talbott, and J.~Zhang, ``{EMOTION}: Expressive motion sequence generation for humanoid robots with in-context learning.''

\bibitem{tonkin_embodiment_2017}
M.~Tonkin, J.~Vitale, S.~Ojha, J.~Clark, S.~Pfeiffer, W.~Judge, X.~Wang, and M.-A. Williams, ``Embodiment, privacy and social robots: May i remember you?'' in \emph{Social Robotics}, A.~Kheddar, E.~Yoshida, S.~S. Ge, K.~Suzuki, J.-J. Cabibihan, F.~Eyssel, and H.~He, Eds.\hskip 1em plus 0.5em minus 0.4em\relax Springer International Publishing, 2017, pp. 506--515.

\bibitem{park_telling_2017}
H.~W. Park, M.~Gelsomini, J.~J. Lee, and C.~Breazeal, ``Telling stories to robots: The effect of backchanneling on a child's storytelling,'' in \emph{2017 12th {ACM}/{IEEE} International Conference on Human-Robot Interaction ({HRI}}, 2017, pp. 100--108.

\bibitem{paplu_harnessing_2022}
S.~Paplu, R.~F. Navarro, and K.~Berns, ``Harnessing long-term memory for personalized human-robot interactions,'' in \emph{2022 {IEEE}-{RAS} 21st International Conference on Humanoid Robots (Humanoids)}, 2022, pp. 377--382.

\bibitem{zhang_towards_2024}
X.~Zhang, R.~Xie, Y.~Lyu, X.~Xin, P.~Ren, M.~Liang, B.~Zhang, Z.~Kang, M.~de~Rijke, and Z.~Ren, ``Towards empathetic conversational recommender systems,'' in \emph{Proceedings of the 18th {ACM} Conference on Recommender Systems}, ser. {RecSys} '24.\hskip 1em plus 0.5em minus 0.4em\relax Association for Computing Machinery, 2024, pp. 84--93.

\bibitem{rashkin_towards_2019}
H.~Rashkin, E.~M. Smith, M.~Li, and Y.-L. Boureau, ``Towards empathetic open-domain conversation models: A new benchmark and dataset,'' in \emph{Proceedings of the 57th Annual Meeting of the Association for Computational Linguistics}, A.~Korhonen, D.~Traum, and L.~M{\`a}rquez, Eds.\hskip 1em plus 0.5em minus 0.4em\relax Florence, Italy: Association for Computational Linguistics, Jul. 2019, pp. 5370--5381.

\bibitem{sharma_computational_2020}
A.~Sharma, A.~S. Miner, D.~C. Atkins, and T.~Althoff, ``A {Computational} {Approach} to {Understanding} {Empathy} {Expressed} in {Text}-{Based} {Mental} {Health} {Support},'' Sep. 2020.

\end{thebibliography}


\newpage
\clearpage
\appendix
\subsection{ChatGPT Prompts}\label{prompts}

OpenAI allows us to include a few different types of prompts in each request: (1) overall context for the conversation, (2) the previous responses given by ChatGPT, (3) instructions from the user to the computer. For each response generation, we passed in the full conversation history for that phase, as well as a context and prompt engineered for the specific phase and turn.

We inputted text from the conversation to personalize the prompts:
\begin{itemize}
    \item $<$name$>$ - the user's preferred name
    \item $<$user's story$>$ - a transcript of the story the user told
    \item $<$story 1, story 2, story 3$>$ - transcripts of the three most emphatically similar stories returned by the intelligent story retrieval model
    \item $<$story selected$>$ - the story that the user most empathized with and chose to reflect on
\end{itemize}

Every prompt also included the following add-on to ensure that the generated responses could be recited by our social robot while still sounding like a natural conversation:
\begin{itemize}
    \item Do not repeat or rephrase what the user says or what you have said, and come up with a playful and creative response (if socially appropriate). You must always ask the user only 1 question and give very concise answers. Only give the words that you would say to the user. Do not refer to the user by name on every turn. Never share links or external media.
\end{itemize}

The contexts and prompts that were made specific to each phase and turn are listed below. The differences between prompting for the \textbf{social agent} and \textbf{social proxy} groups are \textbf{bolded}:

\textsc{\underline{WARMUP}}
\begin{itemize}
    \item \textsc{Context: }You are a social robot named Jibo, and you are getting to know the user.
    \item \textsc{Turn 1: }The given conversation happened the last time you talked. Give only the first thing you would say to start a new conversation based on the conversation from the last time you talked, being specific about the topics discussed in the previous conversation.
    \item \textsc{Turn 2:} \textit{This was left blank to allow the AI to naturally respond to what the user said.}
\end{itemize}

\textsc{\underline{STORY SHARE}}
\begin{itemize}
    \item \textsc{Context: }The user was given a journal that they could write stories in to share with you.
    \item \textsc{Turn 1: }First, respond to the user with a statement. Do not ask a question about what the user said. Then, make sure that you ask the user if they have a story they want to share that they wrote down in their journal or something on their mind they want to tell a story about. Do not ask for a story about what they just talked about in the conversation.
\end{itemize}

\textsc{\underline{STORY RECEIVE (Social Agent Condition)}}
\begin{itemize}
    \item \textsc{Context: }The user just shared the following personal story: $<$user's story$>$. That was a story from the user's life. The following 3 stories are \textbf{personal experiences that you have had in your own life}: $<$story 1, story 2, story 3$>$. You are a robot trying to establish social \textbf{connection with the user based on the stories from your life}.
    \item \textsc{Turn 1: }I am a robot trying to \textbf{connect with you by telling you stories from my life}. First, briefly respond to the user's story in a compassionate manner, making sure to address their emotions in the story. Second, \textbf{tell the user that you have similar experiences to their story from your own life}. Integrate concise 1-2 sentence summaries of all 3 stories into your response. \textbf{All 3 stories are from your life. Make sure to tell the stories as your own.} Finally, direct the user to look at their tablet for full version of the stories from your life.
\end{itemize}

\textsc{\underline{STORY RECEIVE (Social Proxy Condition)}}
\begin{itemize}
    \item \textsc{Context: }The following story is shared by the user $<$user's story$>$. The following 3 stories are \textbf{personal experiences that other people have had}: $<$story 1, story 2, story 3$>$. You are a robot trying to socially \textbf{connect the user to other people}.
    \item \textsc{Turn 1: }I am a robot who \textbf{connects people's stories to the stories of others} in order to improve feelings of social connection. First, briefly respond to the user's story in a compassionate manner, making sure to address their emotions in the story. Second, \textbf{tell the user that you are a robot and that you haven't had experiences like humans do}, but that you love to connect with people, and that\textbf{ many other people can relate to their story}. As examples, introduce \textbf{all 3 of the stories written by other people}. Integrate concise 1 or 2 sentence summaries of all 3 stories into your response. Finally, direct the user to look at their tablet for full version of the stories you mentioned.
\end{itemize}

\textsc{\underline{REFLECTION (Social Agent Condition)}}
\begin{itemize}
    \item \textsc{Context: }The user just shared story 1: $<$user's story$>$. The user read story 2, \textbf{which is from your life:} $<$story selected$>$. You are a robot asking the user to reflect on how they relate to \textbf{the story from your life} (story 2) in relation to their personal story 1 in order to \textbf{improve connection between the user and yourself} through the story they read (story 2). Encourage them to reflect on your story (story 2).
    \item \textsc{Turn 1: }First, tell the user that you are glad that they were able to \textbf{relate to the story from your life} and incorporate a short \textbf{summary of the story they read from your life} (story 2) in your response. Then, point out potential connections between the user's story (story 1) and \textbf{the story from your life} (story 2). Finally, ask the user about \textbf{what they could relate to in your story} (story 2). In your response, do not refer to the stories as 'story 1' and 'story 2'.
    \item \textsc{Turn 2: }Now ask the user how they would feel in \textbf{the situation given in the story from your life} (story 2). \textbf{Incorporate details of the story from your life} (story 2) in your response. In your response, do not refer to the stories as 'Story 1' and 'Story 2'.
    \item \textsc{Turn 3: }Ask the user how they would want their emotions to be addressed if they encountered \textbf{the situation in the story from your life} (story 2), being specific about the situation in \textbf{the story they read from your life} (story 2). In your response, do not refer to the stories as 'Story 1' and 'Story 2'.
    \item \textsc{Turn 4: }Finally, ask the user what they can \textbf{take away from the story from your life} (story 2) and apply to their own experiences. In your response, do not refer to the stories as 'Story 1' and 'Story 2'.
\end{itemize}

\textsc{\underline{REFLECTION (Social Proxy Condition)}}
\begin{itemize}
    \item \textsc{Context: }The user just shared story 1: $<$user's story$>$. The user read story 2, \textbf{which was written by another person:} $<$story selected$>$. You are a robot asking the user to reflect on how they relate to \textbf{the story they read} (story 2) in relation to their personal story 1 in order to \textbf{improve connection between the user and the narrator of the story they read} (story 2). Encourage them to reflect on the story they read (story 2).
    \item \textsc{Turn 1: }First, tell the user that you are glad that they were able to \textbf{relate to the story they read} and incorporate a short \textbf{summary of the story they read} (story 2) in your response. Then, point out potential connections between the user's story (story 1) and \textbf{the story they read} (story 2). Finally, ask the user about \textbf{what they could relate to in the story they read} (story 2). In your response, do not refer to the stories as 'story 1' and 'story 2'.
    \item \textsc{Turn 2:} Now ask the user how they would feel in \textbf{the situation given in the story they read} (story 2). \textbf{Incorporate details of the story they read} (story 2) in your response. In your response, do not refer to the stories as 'Story 1' and 'Story 2'.
    \item \textsc{Turn 3:} Ask the user how they would want their emotions to be addressed if they encountered \textbf{the situation in the story they read} (story 2), being specific about the situation in \textbf{the story they read} (story 2). In your response, do not refer to the stories as 'Story 1' and 'Story 2'.
    \item \textsc{Turn 2:} Finally, ask the user what they can \textbf{take away from the story they read} (story 2) and apply to their own experiences. In your response, do not refer to the stories as 'Story 1' and 'Story 2'.
\end{itemize}

\textsc{\underline{COOL DOWN PHASE}}
\begin{itemize}
    \item \textsc{Context: }You are a robot who is about to end a conversation session with $<$name$>$. You should show $<$name$>$ appreciation for sharing personal stories with you today and tell them that you look forward to seeing them next time.
    \item \textsc{Turn 1: }Incorporate a recap of what happened in the session in your response. Do not repeat or rephrase what the user says or what you have said, and come up with a playful and creative response (if socially appropriate). Only give the words that you would say to the user. Do not refer to the user by name on every turn. Never share links or external media.
\end{itemize}

\subsection{Topics of Stories Retrieved} \label{fig:topics}
\begin{figure*}[h!]
\centering
         \includegraphics[width=\textwidth]{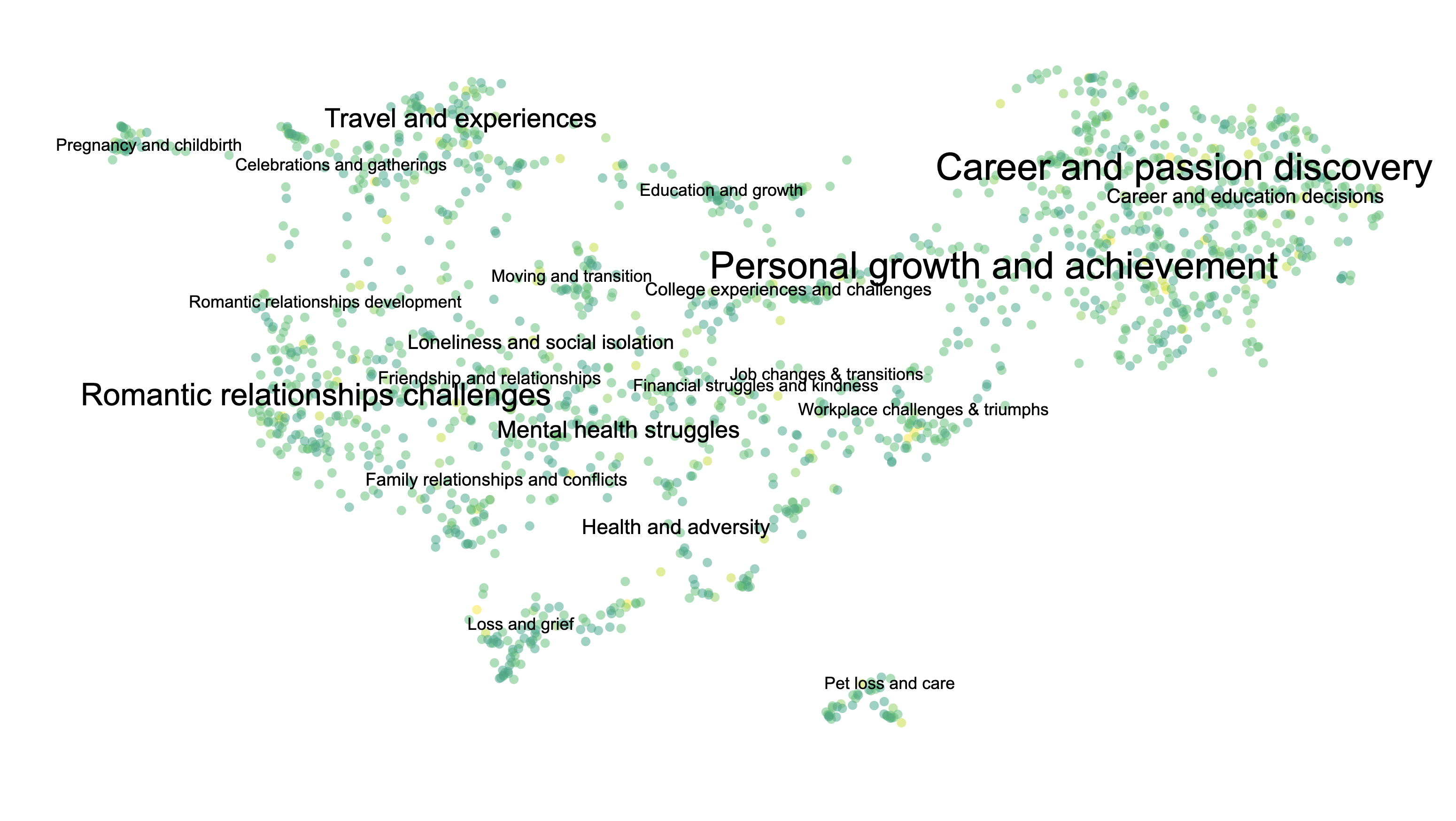}
         \caption{We visualize the topics of stories the robot can share from the \textsc{EmpathicStories} dataset  (obtained with UMAP of ada-002 embeddings). This demonstrates the depth and breadth of stories covering a variety of vulnerable topics such as mental health, relationships, and life changes collected from social media, podcasts, and crowdworkers' stories.
         }
         \label{fig:topicsmap}
\end{figure*}

See Fig.~\ref{fig:topicsmap}.

\subsection{Surveys} \label{surveys}

\subsubsection{Single Item Trait Empathy Scale}
\textit{Rate the extent to which you agree with the following statements.} 

\textit{Strongly disagree (1) to Strongly agree (5)}
\begin{itemize}
    \item I am an empathetic person
\end{itemize}

\subsubsection{Multidimensional Personality Questionnaire - Absorption Scale}
\textit{Rate the extent to which you agree with the following statements.} 

\textit{Strongly disagree (1) to Strongly agree (5)}
\begin{itemize}
    \item I am an empathetic person
    \item While watching a movie, a T.V. show, or a play, I may become so involved that I forgot about myself and my surroundings, and experience the story as if it were real and as if I were taking part in it.
    \item If I wish I can imagine (or daydream) some things so vividly that it’s like watching a good movie or hearing a good story.
    \item I can sometimes recall certain past experiences in my life so clearly and vividly that it is like living them again, or almost so.
    \item If I acted in a play I think I would really feel the emotions of the character and “become” that person for the time being, forgetting both myself and the audience.
\end{itemize}

\subsubsection{Big 5 Personality Test (17)}
\textit{Mark how much you agree with each statement, where each statement starts with \textbf{"I am someone who..."}} 

\textit{Disagree strongly (1) to Agree strongly (5)}
\begin{itemize}
    \item is talkative
    \item tends to find fault with others
    \item does a thorough job
    \item is depressed, blue
    \item is original, comes up with new ideas
    \item is reserved
    \item is helpful and unselfish with others
    \item can be somewhat careless
    \item is relaxed, handles stress well
    \item is curious about many different things
    \item is full of energy
    \item starts quarrels with others
    \item is a reliable worker
    \item can be tense
    \item is ingenious, a deep thinker
    \item generates a lot of enthusiasm
    \item has a forgiving nature
    \item tends to be disorganized
    \item worries a lot
    \item has an active imagination
    \item tends to be quiet
    \item is generally trusting
    \item tends to be lazy
    \item is emotionally stable, not easily upset
    \item is inventive
    \item has an assertive personality
    \item can be cold and aloof
    \item perseveres until the task is finished
    \item can be moody
    \item values artistic, aesthetic experiences
    \item is sometimes shy, inhibited
    \item is considerate and kind to almost everyone
    \item does things efficiently
    \item remains calm in tense situations
    \item prefers work that is routine
    \item is outgoing, sociable
    \item is sometimes rude to others
    \item makes plans and follows through with them
    \item gets nervous easily
    \item likes to reflect, play with ideas
    \item has few artistic interests
    \item likes to cooperate with others
    \item is easily distracted
    \item is sophisticated in art, music, or literature
\end{itemize}

\subsubsection{UBC State Social Connection Survey}

\textit{Please think about how you felt during the past 2 weeks. To what extent do each of the following statements describe how you felt?} 

\textit{Strongly disagree (1) to Strongly agree (7)}
\begin{itemize}
    \item I felt distant from people
    \item I didn’t feel related to most people
    \item I felt like an outsider
    \item I felt like I was able to connect with other people
    \item I felt disconnected from the world around me
    \item I felt close to people
    \item I saw people as friendly and approachable
    \item I felt accepted by others
    \item I had a sense of belonging
    \item I felt a strong bond with other people
\end{itemize}

\subsubsection{Compassionate Love for Humanity Scale}
\textit{Please rate how much you agree with the following statements.} 

\textit{Not at all true of me (1) to Very true of me (7)}
\begin{itemize}
    \item When I hear about someone (a stranger) going through a difficult time, I feel a great deal of compassion for him or her.
    \item It is easy for me to feel the pain (and joy) experienced by others, even though I do not know them.
    \item If I encounter a stranger who needs help, I would do almost anything I could to help him or her.
    \item I feel considerable compassionate love for people from everywhere.
    \item I tend to feel compassion for people even though I do not know them.
    \item One of the activities that provides me with the most meaning to my life is helping others in the world who need help.
    \item I often have tender feelings toward people (strangers) when they seem to be in need.
    \item I feel a selfless caring for most of mankind.
    \item If a person (a stranger) is troubled, I usually feel extreme tenderness and caring.
\end{itemize}


 
\subsubsection{Working Alliance Scale}

\textit{Think about your experience interacting with Jibo, and decide which category best describes your experience.}

\textit{Seldom (1) to Always (5)}
\begin{itemize}
    \item As a result of my interactions with Jibo, I am clearer as to how I might be able to change
    \item What I am doing with Jibo gives me new ways of looking at my problem
    \item I believe Jibo likes me
    \item Jibo and I collaborate on setting goals. 
    \item Jibo and I respect each other 
    \item Jibo and I are working towards mutually agreed upon goals     
    \item I feel that Jibo appreciates me 
    \item Jibo and I agree on what is important for me to work on    
    \item I feel Jibo cares about me even when I do things he doesn't approve of
    \item I feel that the things I do with Jibo will help me accomplish changes I want    
    \item Jibo and I have established a good understanding of the kind of changes that would be good for me
    \item I believe the ways Jibo and I are working with my problems are correct    
\end{itemize}
 
\subsubsection{Robot Perceived Empathy Scale}

\textit{Thinking about your interactions with Jibo, rate your agreement with the following statements}

\textit{Strongly disagree (1) to Strongly agree (5)}
\begin{itemize}
    \item Jibo appreciates exactly how the things I experience feel to me.
    \item Jibo knows me and my needs.
    \item Jibo cares about my feelings.
    \item Jibo does not understand me.    
    \item Jibo perceives and accepts my individual characteristics.
    \item Jibo usually understands the whole of what I mean.    
    \item Jibo reacts to my words but does not see the way I feel.
    \item Jibo seems to feel bad when I am sad or disappointed.    
    \item Whether thoughts or feelings I express are “good” or “bad” makes no difference to Jibo's actions toward me.
    \item No matter what I tell about myself, Jibo acts just the same.    
    \item Jibo comforts me when I am upset.
    \item Jibo encourages me.    
    \item Jibo praises me when I have done something well.
    \item Jibo helps me when I need it.   
    \item Jibo knows when I want to talk and lets me do so.
    \item Jibo‘s response to me is so fixed and automatic that I do not get through to it.    
    \item The way Jibo acts feels natural.
    \item Jibo knows what it is doing.    
    \item Jibo is responsible for its actions.
    \item When I interact with Jibo, I feel anxious.    
\end{itemize}
 
\subsubsection{Dyadic Trust Scale}

\textit{Again, thinking about your relationship with Jibo, rate your agreement with the following statements.}

\textit{Strongly disagree (1) to Strongly agree (5)}
\begin{itemize}
    \item Jibo is primarily interested in his own welfare
    \item There are times when Jibo cannot be trusted
    \item Jibo is perfectly honest and truthful with me
    \item I feel that I can trust Jibo completely
    \item Jibo is truly sincere in his promises
    \item I feel that Jibo does not show me enough consideration   
    \item Jibo treats me fairly and justly 
    \item I feel that Jibo can be counted on to help me    
\end{itemize}
 
\subsubsection{Godspeed Surveys}

\textit{Rate your impression of Jibo on the following scales (1 to 5)}
\begin{itemize}
    \item Fake vs. Natural
    \item Machinelike vs. Humanlike
    \item Unconscious vs. Conscious
    \item Artificial vs. Lifelike
    \item Moving rigidly vs. Moving elegantly
    \item Dead vs. ALive
    \item Stagnant vs. Lively
    \item Mechanical vs. Organic
    \item Artificial vs. Lifelike
    \item Inert vs. Interactive
    \item Apathetic vs. Responsive
    \item Dislike vs. Like
    \item Unfriendly vs. Friendly
    \item Unkind vs. Kind
    \item Unpleasant vs. Pleasant
    \item Awful vs. Nice
    \item Incompetent vs. Competent
    \item Ignorant vs. Knowledgeable
    \item Irresponsible vs. Responsible
    \item Unintelligent vs. Intelligent
    \item Foolish vs. Sensible
    \item Anxious vs. Relaxed
    \item Calm vs. Agitated
    \item Still vs. Surprised
\end{itemize}

\subsubsection{Spitale et al. Survey}

\textit{Thinking about your interactions with Jibo, rate your agreement to the following statements}

\textit{Strongly disagree (1) to Strongly agree (5)}
\begin{itemize}
    \item If I were worried, Jibo would make me feel better
    \item If I were nervous, Jibo would make me feel more calm
    \item If I were upset, Jibo would make me feel better
    \item I had fun listening to Jibo     
    \item Jibo made me laugh
    \item I enjoyed talking to Jibo
    \item It was nice being with Jibo
    \item I found Jibo easy to interact with   
    \item I think I can interact with Jibo without any help 
    \item I think I can interact with Jibo when there is someone around to help me
    \item I think I can interact with Jibo when I have good instructions
    \item I enjoy Jibo talking with me    
    \item I found Jibo enjoyable
    \item I found Jibo fascinating
    \item I consider Jibo a nice partner to talk to
    \item I find Jibo nice to interact with    
    \item I feel Jibo understands me
    \item I think Jibo is nice
    \item When interacting with Jibo I felt like I was talking to a real person
    \item It sometimes felt as if Jibo was really looking at me    
    \item I can imagine Jibo to be a living creature
    \item I often think Jibo is a real person
    \item Sometimes Jibo seems to have real feelings
    \item I would trust Jibo if it gave me advice
    \item I would follow the advice Jibo gives me    
\end{itemize}

\subsection{Participant Materials} \label{materials}
\begin{figure}[h]
  \includegraphics[width=\linewidth]{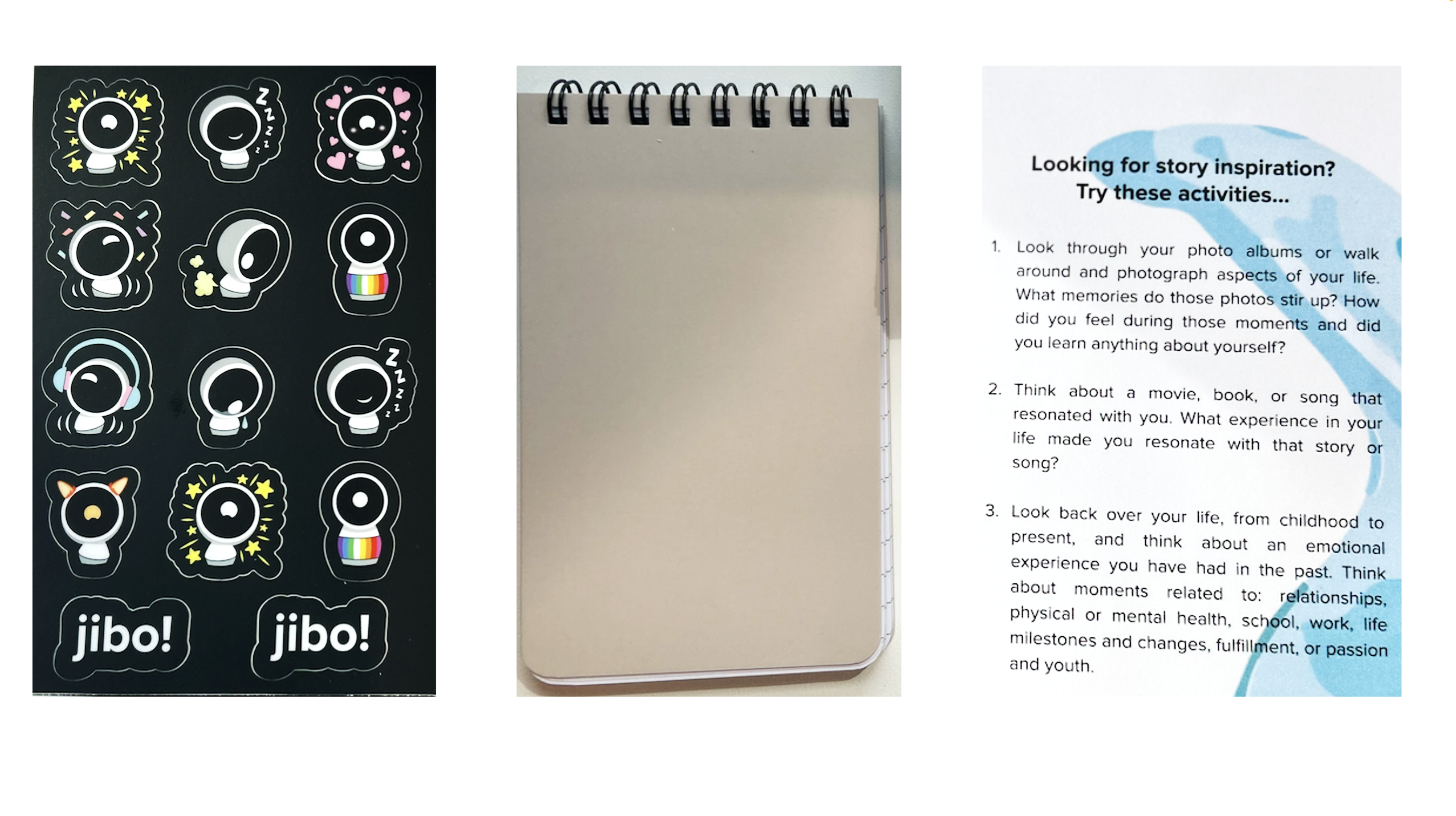}
  \caption{Study materials provided to participants}
  \label{fig:studymaterials}
\end{figure}
\subsection{Qualitative Codebook}

\begin{table}[h!]
\centering
\footnotesize
\caption{Full codebook with definitions}
\label{tab:code_definitions}
\resizebox{\linewidth}{!}{ 
\begin{tabularx}{\linewidth}{p{0.25\linewidth}X} 
\hline
\textbf{Code} & \textbf{Definition} \\
\hline
identifying connections & the robot helps draw connections between the user's life and other people's stories  \\
\hline
connection beyond stories & find the robot as a catalyst in helping them connect with others outside of the stories shared  \\
\hline
empathy and perspective taking & the robot helps the participant fully absorb into another person's story, or see the story from another person's shoes  \\
\hline
diverse perspectives & the stories the robot shares offer different perspectives on situations discussed  \\
\hline
judgement-free & finds it easy to talk to the robot because it cannot judge them / shared things with the robot that could not be shared with other people  \\
\hline
reframing & finds the robot a helpful impartial third-party agent to reframe thoughts and experiences  \\
\hline
active listening & feels that the robot listens and understands them, mentions that the robot remembers information from previous sessions or reiterates what they say  \\
\hline
counseling & mentions appreciation for the robot's advice, helping them work through situations or emotions  \\
\hline
encouraging & finds that the robot offered reassurance or validation during conversation  \\
\hline
question authenticity & mentions that the stories or interaction with the robot is artificial in some manner  \\
\hline
disagreement & mentions disagreeing with the story or understanding the story but not empathizing  \\
\hline
repetition & mentions that the robot unnaturally repeats stories or questions  \\
\hline
mechanical & mentions that the robot is too technical or unfeeling, whether by speech or physical design  \\
\hline
security & mentions discomfort about the robot or robots in general infringing on their security and privacy  \\
\hline
\end{tabularx}
}
\end{table}


\subsection{Supplementary Results from Survey Analysis} \label{supplementarysurveyanalysis}

\begin{figure}[t]
\centering
  \includegraphics[width=\linewidth]{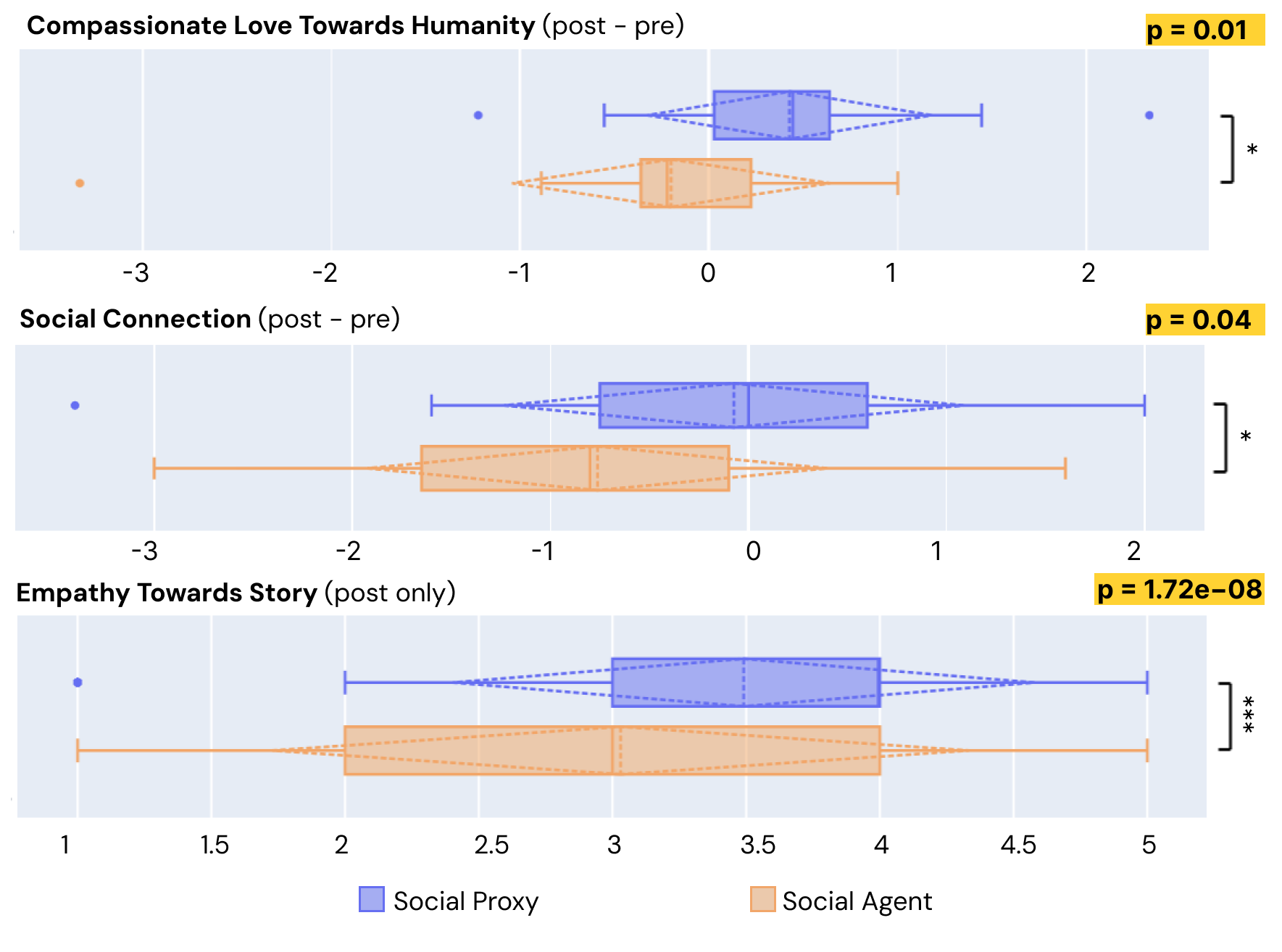}
  \caption{\highlight{Effects on change in compassionate love to humanity and state connection. changes in Compassionate Love Towards Humanity and Social Connection, as well as Empathy Towards the Story were significantly higher in the \textbf{social proxy} vs. \textbf{social agent} condition. Note p-values are obtained from Mann-Whitney u-test.}}
  \vspace{-5pt}
  \label{fig:pyschometrics}
\end{figure}
\subsubsection{Comparing Social Proxies and Social Agents}

First, looking at changes in social connection before and after the study within each condition, we find that compassionate love for humanity increases and state social connection decreases slightly from the pre-study survey to the post-study survey in the \textbf{social proxy} condition, although not statistically significantly so for either measure. 
Changes from the pre-study survey to the post-study survey ($post - pre$) for compassionate love for humanity and state social connection across the conditions are shown in Fig.~\ref{fig:pyschometrics}. Participants in the \textbf{social proxy} condition, where the robot made clear that the stories came from other people, had significantly higher changes in state social connection ($u = 276.0, p=0.039, d = 0.38$/medium effect size) and compassionate love for humanity ($u = 294.5, p=0.01, d = 0.48$/large effect size) compared to the  \textbf{social agent} condition, where the robot shares the stories as its own, as shown by the Mann-Whitney u-test between the pre and post study surveys. 
Participants in the \textbf{social proxy} condition had statistically significantly higher empathy towards the stories they read compared to the  \textbf{social agent} condition  ($u = 192853.5, p =1.72e-08, d = 0.20$/small effect size), as we can see from Fig.~\ref{fig:pyschometrics}. 


Interestingly, while no significant changes were observed over time in the \textbf{social proxy} condition, in the \textbf{social agent} condition, state social connection significantly decreased from the start to the end of the study ($u = 362.5$, $p = 0.012$, $d = 0.44$/medium effect size). 


\subsubsection{Confounding Variables and Statistical Validity}\label{structuralanalysis}


\begin{figure}[t]
  \includegraphics[width=\linewidth]{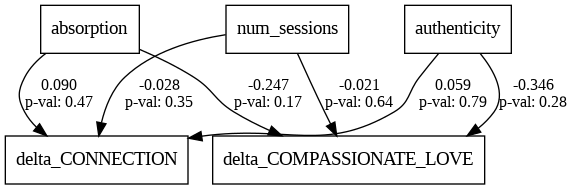}
  \caption{\highlight{Structural analysis of variables related to story authenticity and their effect on change in connection (delta\_CONNECTION) and change in compassionate love towards humanity (delta\_COMPASSIONATE\_LOVE)}}
  \label{fig:structural}
\end{figure}
Because the robot is automatically retrieving empathetically similar stories and dialogue is generated using ChatGPT, there is the potential contribution of the perceived authenticity of the robot's story genuinely coming from the robot's experiences. We ran numerous experiments with ChatGPT to try to convert stories into ``robot-believable'' form, but found that ChatGPT was not capable of this task (stories were either altered too much or the prompt was ignored). Thus, we conduct post-hoc structural analyses on the number of sessions, trait absorption, and perceived authenticity of the stories, as the number of sessions could be related to perceived authenticity and trait absorption is related to the extent to which they can believe or absorb themselves in an unrealistic scenario.

We report results of our structural analysis to strengthen the point that  quantitative results are not heavily influenced by potential confounds, including user's trait absorption, total number of sessions participants did with the robot, and mentions of whether or not the robot's stories were authentic or not (as stated in the participant interviews). In Figure \ref{fig:structural}, we show that these variables do not significantly impact the outcome variables of change in connection and change in compassionate love towards humanity (0.17 $<$ p $<$ 0.79). Thus, we argue that the changes we observe are most attributed to the manipulation across conditions.

\subsubsection{Discussion of Supplementary Findings}
We found that participants showed more significant increases in compassionate love towards humanity, state social connection, and empathy towards stories when the robot delivered stories as a social proxy, rather than delivering the stories as its own experiences. 

We did not find any significant decrease or increase in social connection before and after the study in the \textbf{social proxy} condition. However, we observed a significant drop in social connection before and after the study for the \textbf{social agent} condition. These results could potentially be explained by the participants' negative reactions to stories shared as the robot's own experience, which conflates with transparency, and users could discredit empathy they might have towards the narrator. This explanation is in line with prior work that show participants have a strong preference for the authenticity of a story origin \cite{bhattacharjee_i_2022, shen_empathy_2024} -- In particular, knowing or being transparent that a story is written by a person allows a reader to feel greater connection.

Finally, we did not observe significant changes pre and post and connection and our structural analysis reflects what external factors did not influence this result, but we acknowledge that a limitation of our work is not probing deeper through the interviews about other external factors related to their change in connection. We also hypothesize that more sessions might be needed to see a consistent change in social connection. From this limitation, future works for longitudinal studies similar to our work should carefully design interview questions and methods for gathering a deeper contextual understanding of user experiences, considering both internal and external factors of the study.

\end{document}